\documentclass[a4paper, 10pt]{article}
\usepackage{jheppub} 
\usepackage{hyperref}
\usepackage[utf8x]{inputenc} 
\usepackage[dvipsnames]{xcolor}
\usepackage{booktabs}
\usepackage{arydshln}
\usepackage{gensymb}

\DeclareMathAlphabet{\mathsfit}{\encodingdefault}{\sfdefault}{m}{sl}

\definecolor{darkred}{rgb}{0.6, 0, 0}
\definecolor{darkblue}{rgb}{0, 0, 0.6}
\hypersetup{colorlinks=true, linkcolor=darkblue, urlcolor=darkblue, citecolor=darkred}

\makeatletter
\newcommand{\@fpheader}{}   % do nothing
\makeatother

\title{Boosting probes of ${\cal CP}$ violation in the  top Yukawa coupling with Deep Learning}

\author[a]{Waleed Esmail,}
\author[b]{\! A. Hammad,}
\author[c,d]{\! Adil Jueid}
\author[e,f]{\! and Stefano Moretti}

\affiliation[a]{Institut f\"ur Kernphysik, Universit\"at M\"unster, \\ Wilhelm-Klemm-Str. 9, 48149 M\"unster, Germany}
\affiliation[b]{Theory Center, IPNS, KEK,  1-1 Oho, Tsukuba, Ibaraki 305-0801, Japan}
\affiliation[c]{Particle Theory and Cosmology Group, Center for Theoretical Physics of the Universe, \\
Institute for Basic Science (IBS), \\ Daejeon, 34126, Republic of Korea}
\affiliation[d]{Cosmology, Gravitation and Astroparticle Physics Group,  
Center for Theoretical Physics of the Universe, \\ Institute for Basic Science (IBS), \\ Daejeon, 34126, Republic of Korea}
\affiliation[e]{School of Physics $\&$ Astronomy, University of Southampton, \\ Highfield, Southampton SO17 1BJ, U.K.}
\affiliation[f]{Department of Physics $\&$  Astronomy, Uppsala University, \\ Box 516, 75120 Uppsala, Sweden}

\begin{document}

\abstract{The precise measurement of the top-Higgs coupling is crucial in particle physics, offering insights into potential new physics Beyond the Standard Model (BSM) carrying  ${\cal CP}$ Violation (CPV) effects. In this paper, we explore the  ${\cal CP}$ properties of a Higgs boson coupling with a top quark pair, focusing on events where the Higgs state decays into a pair of $b$-quarks and the top-antitop system  decays leptonically. The novelty of our analysis resides in the exploitation of two conditional Deep Learning (DL) networks: a  Multi-Layer Perceptron (MLP) and a  Graph Convolution Network (GCN). These models are trained for selected CPV phase values and then used to interpolate all possible values ranging from $0$ to  $\pi/2$. This enables a comprehensive assessment of sensitivity across all ${\cal CP}$ phase values, thereby streamlining the process as the models are trained only once. Notably, the conditional GCN exhibits superior performance over the conditional MLP, owing to the nature of graph-based Neural Network (NN) structures. Specifically, for Higgs top coupling modifier set to 1, with $\sqrt{s}= 13.6$ TeV and integrated luminosity of $3$ ab$^{-1}$  GCN excludes the ${\cal CP}$ phase larger than $5^\circ$ at $95.4\%$ Confidence Level (C.L).  Our Machine Learning (ML) informed findings indicate that assessment of the ${\cal CP}$ properties of the Higgs coupling to the $t\bar t$ pair can be within reach of the High Luminosity Large Hadron Collider (HL-LHC), quantitatively surpassing the sensitivity of more traditional approaches. }

\maketitle
%\tableofcontents

\section{Introduction}
\label{S:Intro}
Since when first discovered in  
 the long-lived $K$-meson rare decay channel $K_L\rightarrow2\pi$ back in 1964 \cite{Christenson:1964fg},  CPV has attracted conspicuous theoretical
 interest, given that studying its dynamics in laboratory experiments may eventually open up a window of understanding on the matter-antimatter asymmetry in the Universe. Unfortunately, all phenomena of the first kind (including those measured also in the $D$- and 
$B$-meson sectors) can be explained using the  Kobayashi-Maskawa mechanism \cite{Kobayashi:1973fv},
which, while representing a success of the SM, is, however, not enough to explain the latter \cite{Cohen:1991iu,Cohen:1993nk,Morrissey:2012db}. Over the past few decades, many Beyond the BSM scenarios that can accommodate additional CPV sources, whether spontaneous or explicit,  have been proposed so as to remedy such a SM flaw. While theoretically  viable, these are all strongly constrained by experiments. In particular, very precise measurements of the Electric Dipole Moments (EDMs) of, e.g.,  electron and neutron, \cite{Andreev:2018ayy,Cairncross:2017fip,Abel:2020gbr} have already
placed severe limits on many new CPV sources \cite{Yamanaka:2017mef,Safronova:2017xyt,Chupp:2017rkp}. In fact, the sensitivities attained herein are far above the SM predictions \cite{Pospelov:2005pr,Yamaguchi:2020eub,Yamaguchi:2020dsy},
yet, EDM measurements, both those above and others, being very inclusive in their nature, are  unlikely to determine the actual interactions affected by CPV. Conversely, collider experiments, despite having weaker sensitivities to CPV  effects in comparison, can afford one, thanks to the huge variety of exclusive quantities  that one can define in such settings, with an insight into the actual CPV dynamics. 

This is particularly true for BSM frameworks with extended Higgs sectors \cite{Bento:1991ez,Lee:1973iz,Lee:1974jb,Branco:2011iw,Weinberg:1976hu}, wherein non-zero EDMs are expected to be the first signal of CPV with collider effects instead being able to provide additional information on it, see, e.g., Refs.
\cite{ElKaffas:2006gdt,Berge:2008wi,Shu:2013uua,Mao:2014oya,Chen:2015gaa,Keus:2015hva,Fontes:2015mea,Bian:2016awe,Mao:2016jor,Hagiwara:2016zqz,Chen:2017com,Cao:2020hhb,Azevedo:2020fdl,Azevedo:2020vfw,Antusch:2020ngh,Cheung:2020ugr,Kanemura:2021atq,Low:2020iua}. Indeed, since the discovery of the 125 GeV Higgs boson at the LHC in 2012 \cite{Aad:2012tfa,Chatrchyan:2012xdj,Aad:2015zhl}, testing its ${\cal CP}$ properties has been high on the ATLAS and CMS agendas, as the SM has a definite prediction in this respect, {\it i.e.}, $0^+$\footnote{This is only partly true as the Higgs boson of the SM receives some ${\cal CP}$ violation via CKM mixing through loop corrections which is however quite small.}, so that any deviations from this would be a signal of BSM physics. At present, the status of such measurements is that they 
are all consistent with the CP-even (or scalar) state
of the SM, yet, the possibility of a CP-odd (or pseudoscalar) component to it ({\it e.g.}, through mixing with another Higgs state) cannot be definitely excluded.

A simple BSM setup, exploiting the fact that Nature appears to privilege doublet representations of a Higgs field, is the 2-Higgs Doublet Model (2HDM) \cite{Branco:2011iw}.%, which we will adopt as illustrative example  in this paper.
Herein, an effective method to test CPV effects is to study the Yukawa interactions between any Higgs boson (the SM-like one or others) and the top (anti)quark via the Lagrangian term
\begin{equation}
-\mathcal{L}_{tt\phi_0}\propto \bar{t}\left(g_S+\textrm{i}g_P\gamma^5\right)t\phi_0 ,
\end{equation}
where $\phi_0$ refers to a generic Higgs state with mixed ${\cal CP}$ quantum numbers and $g_{S(P)}$ refers to its corresponding scalar(pseudoscalar) coupling to a $t\bar t$ pair.
This vertex is   of importance because a top (anti)quark decays quickly enough so that the information emerging from such an interaction feeds into its final state distributions. Specifically, the ${\cal CP}$ quantum 
numbers (and also spin) of the Higgs state can  be accessed through these, albeit on a statistical basis. 

Phenomenologically, there are a lot of studies in the literature trying to 
test CPV in $t\bar{t}\phi_0$ interactions at colliders \cite{Schmidt:1992et,Mahlon:1995zn,Asakawa:2003dh,BhupalDev:2007ftb,He:2014xla,Boudjema:2015nda,Mileo:2016mxg,Buckley:2015vsa,AmorDosSantos:2017ayi,Azevedo:2017qiz,Bernreuther:2017cyi,Hagiwara:2017ban,Ma:2018ott,Cepeda:2019klc,Faroughy:2019ird,Cheung:2020ugr,Bahl:2020wee,Cao:2020hhb,Azevedo:2020fdl,Azevedo:2020vfw,Bahl:2021dnc,Barman:2021yfh,Bahl:2022yrs,Bahl:2023qwk} (see also Ref. \cite{Hammad:2025ewr} for a short review). It is the purpose of this paper the one of contributing to the  endeavour of extracting CPV effects from $pp\to t\bar{t}\phi_0$ at the HL-LHC \cite{Gianotti:2002xx} through a novel approach exploiting two alternative DL methods: a conditional MLP and a conditional GCN.   

The paper is organised as follows. We describe our theoretical setup in Sec. \ref{sec:model}, including its phenomenological manifestations through the $pp\to t\bar t \phi_0$ process, while in Sec. \ref{sec:observables} we introduce the kinematical observables that we will exploit in  our   numerical analysis. Sec. \ref{sec:DNNs} is devoted to describe the aforementioned conditional Deep NNs (DNNs)  in some detail. We then perform the DL analyses in Sec. \ref{sec:DL} and produce our final results in Sec. \ref{sec:results}. We then conclude in Sec. \ref{sec:summary}. (There is also an appendix where we test the DNN activities in terms of a toy example.)

\begin{figure}[!t]
\centering
\vspace{-1cm}
\includegraphics[width=0.89\linewidth]{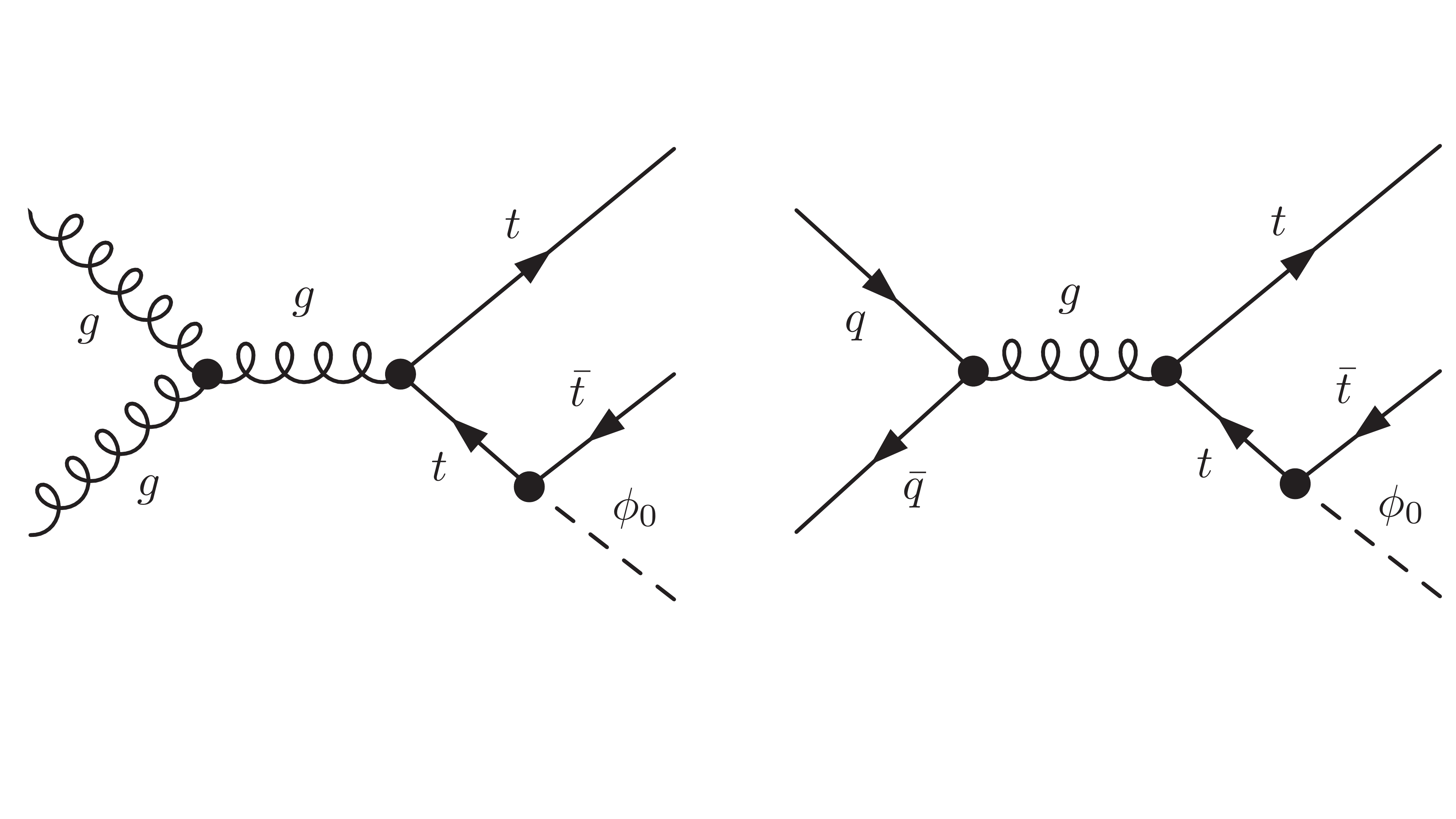}
\vspace{-1.5cm}
\caption{Representative Feynman diagrams corresponding to $t\bar{t}\phi_0$ production at Leading Order. Here we show the production through $gg$ fusion (left) and $q\bar{q}$ annihilation (right).}
\label{fig:diagrams}
\end{figure}

%%%%%%%%%%%%%%%%%%%%%%%%%%%%%%%%
\section{Setup}
\label{sec:model}
%%%%%%%%%%%%%%%%%%%%%%%%%%%%%%%%

%%%%%%%%%%%%%%%%%%%%%%%%%%%%%%%
\subsection{The $pp\to t\bar{t}\phi_{0}$ process}
%%%%%%%%%%%%%%%%%%%%%%%%%%%%%%%

We consider the production of a scalar boson ($\phi_0$) with a mass of $m_{\phi_0} = 125$ GeV at the LHC in association with a top quark pair. We parameterize the $t\bar{t}\phi_0$ Lagrangian as follows:
\begin{eqnarray}
    - {\cal L}_{t\bar{t}\phi_0} = \frac{m_t}{\upsilon} \kappa_{tt} \bar{t} \bigg(\cos\theta_{tt} + i \gamma_5 \sin\theta_{tt} \bigg) t \phi_0 ,
\end{eqnarray}
where $\kappa_{tt}$ and $\theta_{tt}$, assumed to be real valued, are the free parameters of this simplified model of Yukawa interaction. (A pure SM Higgs boson is realised for $\kappa_{tt} = 1$ and $\theta_{tt} = 0 \degree$). We assume instead that the coupling modifiers of the Higgs boson to the other fermions and gauge bosons are SM-like while we allow for the same degree of  CPV to  be transferred from $\phi_0$ to its decay products. Therefore, the Lagrangian for the interaction of $\phi_0$ with the other fermions ($f=d,u,s,c,b$ and $\ell$, with $\ell=e,\mu,\tau$) and gauge bosons ($V=W^\pm,Z$) of the SM is parameterized as
\begin{eqnarray}
    {\cal L} = - \sum_f \frac{m_f}{\upsilon} \bar{f} \bigg(\cos\theta_{ff} + i \gamma_5 \sin\theta_{ff} \bigg) f \phi_0 + \sum_{V} \frac{2 m_V^2}{\upsilon} \cos\theta_{VV} V^\mu V_\mu \phi_0,
\end{eqnarray}
where we assume that $\theta_{ff} = \theta_{VV} = \theta_{tt}$ in our analysis. The master formula for the production of $t\bar{t}\phi_0$ final states at a $pp$ collider is given by
\begin{eqnarray}
    \sigma(pp \to t\bar{t}\phi_0) \equiv \sum_{i,j} \int \mathrm{d}x_i \mathrm{d}x_j f_{i/p}(x_i, \mu_F^2) f_{j/p}(x_j,\mu_F^2) \hat{\sigma}_{ij \to t \bar{t}\phi_0}(\hat{s}; \mu_R^2, \mu_F^2),
\end{eqnarray}
where $f_{i/p}(x_i, \mu_F^2)$ is the probability for parton $i$ to carry a momentum fraction $x_i = p_i/P$ of the proton momentum $P$ at a scale $\mu_F$ (factorization scale) and $\hat{\sigma}_{ij\to t\bar{t}\phi_0}$ is the partonic cross section evaluated at the scales $\mu_R$ (renormalization scale) and $\mu_F$. Representative Leading Order (LO) Feynman diagrams for such a process are shown in Fig. \ref{fig:diagrams}. The scattering amplitude receives two contributions at LO: $gg$ fusion (left panel of Fig. \ref{fig:diagrams}) and $q\bar{q}$ annihilation (right panel of Fig. \ref{fig:diagrams}).
The gluon fusion contribution is important for small momentum fractions while the (anti)quark contribution is dominant at moderate and large momentum fractions. However, the total cross section is dominated by the contribution of gluons when integrated over all the momentum fractions (about $58\%$ for $\theta_{tt} = 0\degree$).

\begin{figure}[!t]
    \centering
    \includegraphics[width=0.55\linewidth]{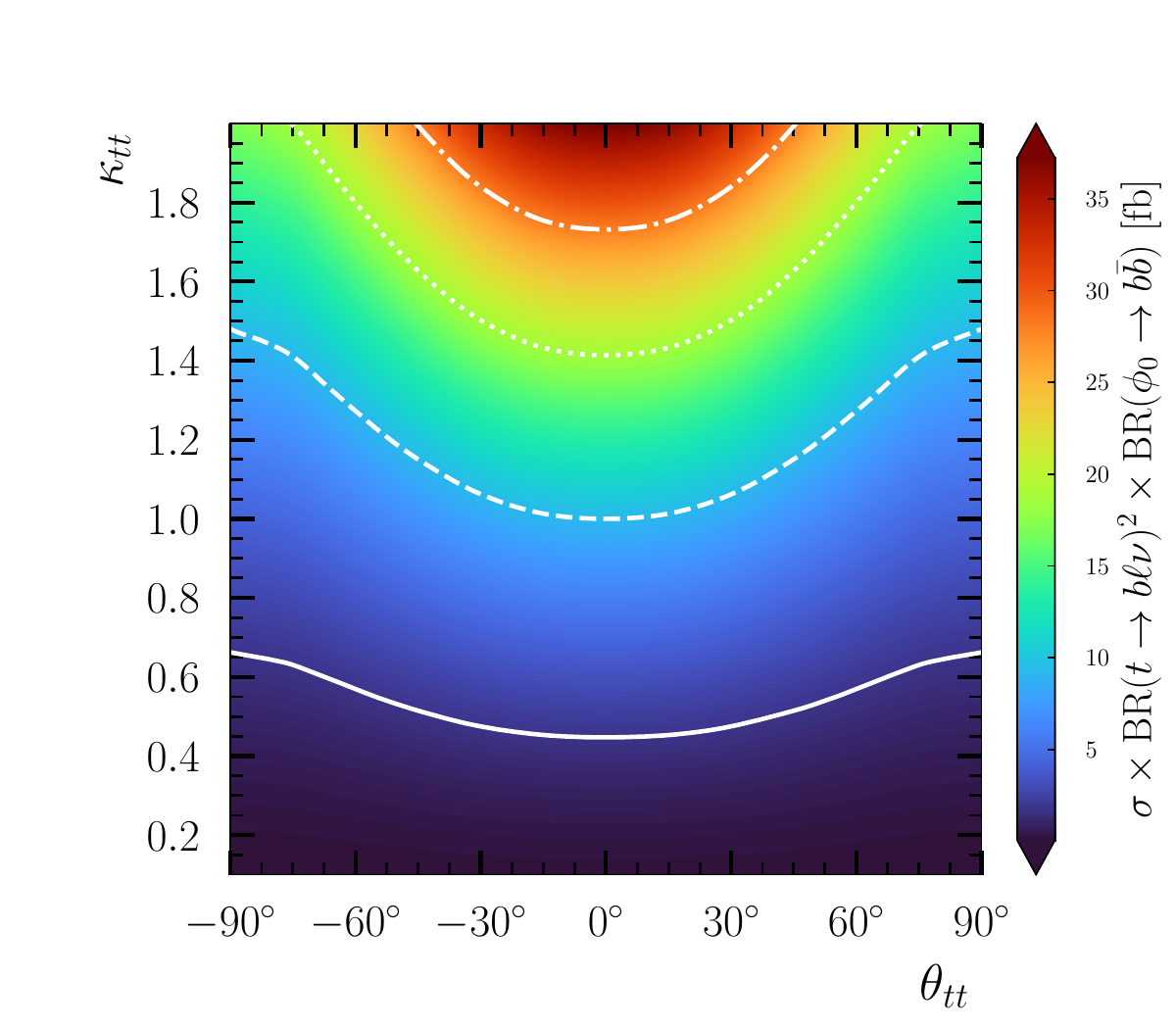}
    \caption{Production cross section of $t\bar{t}\phi_0$ times the product of the Branching Ratios (BRs) assuming the top quark decaying leptonically and the scalar boson decaying into $b\bar{b}$ projected on the plane of $\kappa_{tt}$ and $\theta_{tt}$. We also show the contours corresponding to $\sigma/\sigma_{\rm SM} = 0.2, 1, 2, 3$ in solid, dashed, dotted and dash-dotted white lines, respectively. Here, $\sigma_{\rm SM}$ is the cross section for $\kappa_{tt} = 1$ and $\theta_{tt} = 0\degree$.}
    \label{fig:crosssection}
\end{figure}

We employ \textsf{Madgraph5\_aMC@NLO} version 3.4.1 \cite{Alwall:2011uj, Alwall:2014hca} to calculate the production cross section and using the LO PDF set \textsf{NNPDF40\_lo\_as\_01180}  \cite{NNPDF:2021njg} through the \textsf{LHAPDF} library \cite{Buckley:2014ana}. As for the renormalization and factorization scales, we have adopted a common (dynamical) choice for the central theory predictions, {\it i.e.},  
\begin{eqnarray}
    \mu_F^0 = \mu_R^0 = \frac{1}{2} \sum_i \sqrt{p_{T,i}^2 + m_i^2},
\end{eqnarray}
which is the sum of the transverse masses of all the final-state particles divided by two. We note that the theory uncertainties at LO are dominated by those arising from scale variation ($\simeq 20\%$--$30\%$) while PDF uncertainties are negligible ($\simeq 2\%$--$4\%$). The $K$ factor measuring the size of QCD Next-to-LO (NLO) corrections  can be of order $\approx 1.1$--$1.2$  (see Refs. \cite{Frederix:2011zi,Maltoni:2016yxb} for comprehensive analyses of QCD corrections), but we do not take it into account in our simulations for consistency reasons, as not all our backgrounds are known to NLO accuracy. In Fig. \ref{fig:crosssection} we display the cross section defined as 
\begin{eqnarray}
    \sigma \equiv \sigma(pp \to t\bar{t}\phi_0) \times {\rm BR}(t\to b\ell\nu)^2 \times {\rm BR}(\phi_0 \to b\bar{b}),\qquad \ell=e,\mu,
\end{eqnarray}
over the plane of $\theta_{tt}$ and $\kappa_{tt}$. The partial width of $\phi_0 \to b\bar{b}$ is calculated at Next-to-next NLO (N$^3$LO) using a full resummed running $b$-quark mass \cite{Djouadi:2005gi,Djouadi:2005gj,Chetyrkin:1996sr,Chetyrkin:1997vj}. We can see that the cross section varies between $2$ and $36$ fb with the maximum being for large $\kappa_{tt}$ and $\theta_{tt} \approx 0$.  Notice that, given that we wish to study (charge and spin) correlations in the $t\bar{t}$ system, we are considering here fully leptonic decays of it.

%%%%%%%%%%%%%%%%%%%%%%%%%%%%%%%%%%%%%%%%%%%%%%%%%%%
\subsection{Mass reconstructions of top (anti)quarks}
\label{sec:reconstruction}
%%%%%%%%%%%%%%%%%%%%%%%%%%%%%%%%%%%%%%%%%%%%%%%%%%%

\begin{figure}[!t]
    \centering
    \includegraphics[width=0.49\linewidth]{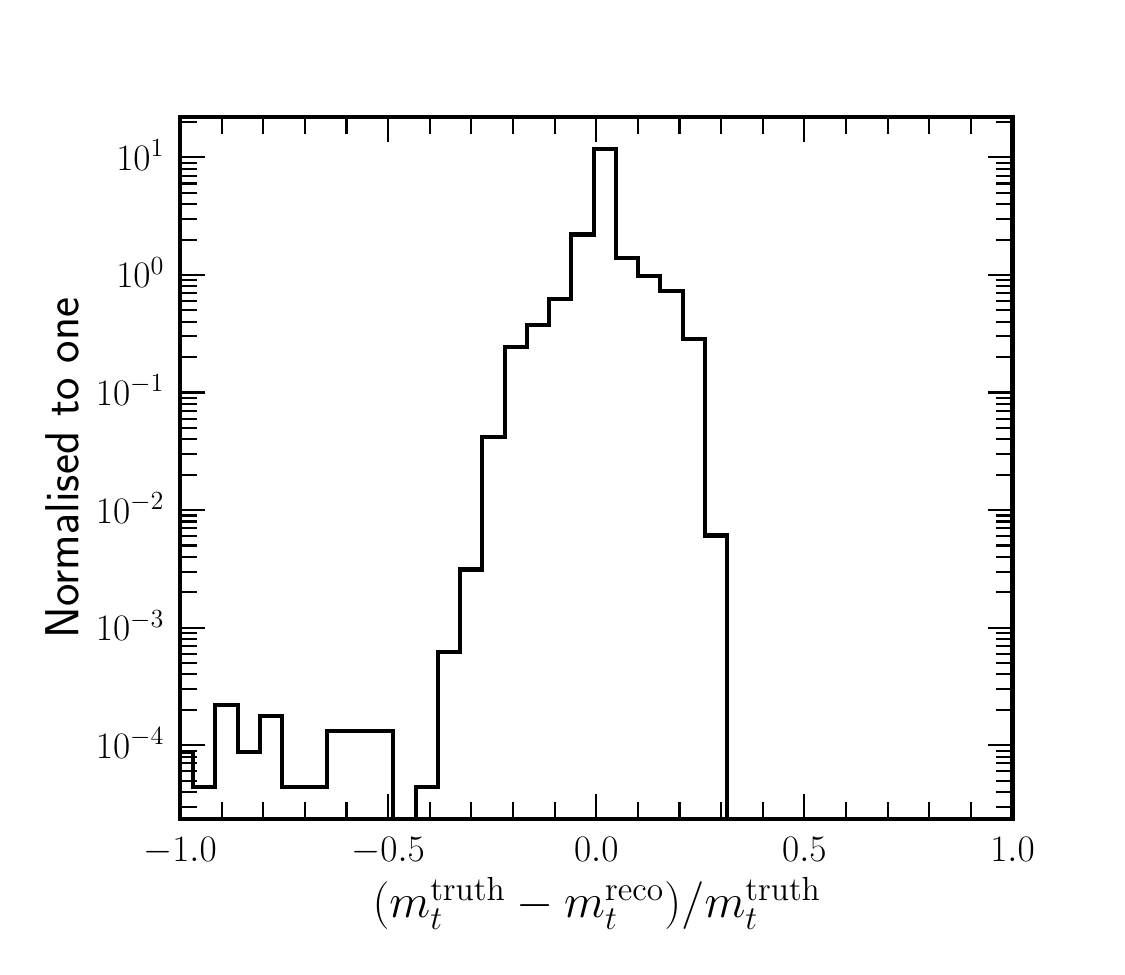}
    \hfill
    \includegraphics[width=0.49\linewidth]{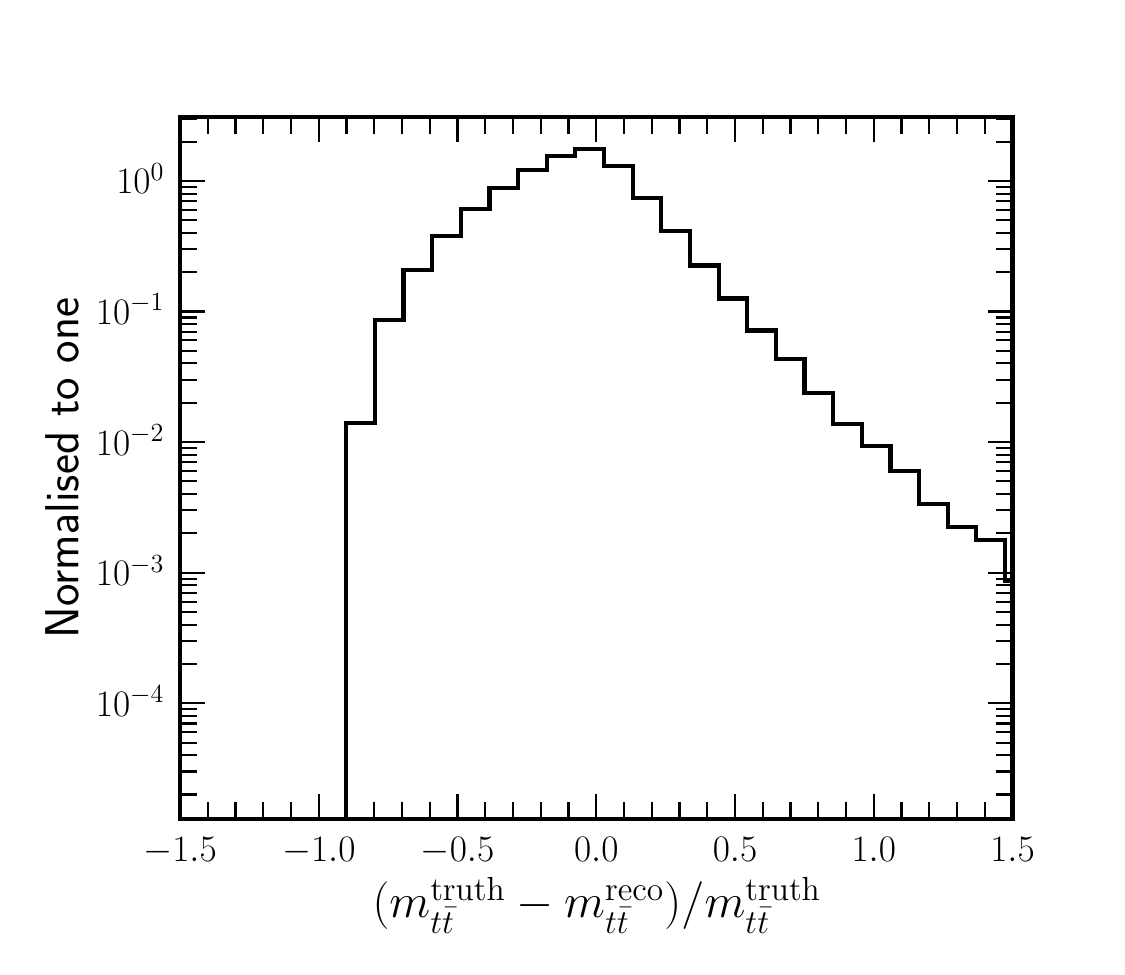}
    \caption{The $m_t$ (left) and $m_{t\bar{t}}$ (right) resolution of the analytical mass reconstruction method outlined in the text. Here, $m_t^{\rm truth}$ and $m_{t\bar{t}}^{\rm thruth}$ refer to the invariant mass of the (anti)top and the $t\bar{t}$ system in the last parton shower step before the decay of the (anti)top quark.}
    \label{fig:tt:resolution}
\end{figure}

The dileptonic decays of the $t\bar{t}$ system lead to two neutrinos in the final state implying an ambiguity in the reconstruction of the (anti)top invariant mass. The reconstruction of the full $t\bar{t}$ invariant mass is also very important to construct the ${\cal CP}$-sensitive observables described in Sec. \ref{sec:observables}. At hadron colliders, the only handle to neutrinos in this process is through the total missing momentum given as the longitudinal momentum of the initial partons is unknown. The conservation of the total momentum of the (anti)top quark and Higgs boson leads to the following constraints:
\begin{eqnarray}
    M_{W^+}^2 &=& (E_{\ell^+} + E_\nu)^2 - (p_{\ell^+}^x + p_\nu^x)^2 - (p_{\ell^+}^y + p_\nu^y)^2 - (p_{\ell^+}^z + p_\nu^z)^2, \nonumber \\ 
    M_{W^-}^2 &=& (E_{\ell^-} + E_{\bar{\nu}})^2 - (p_{\ell^-}^x + p_{\bar{\nu}}^x)^2 - (p_{\ell^-}^y + p_{\bar{\nu}}^y)^2 - (p_{\ell^-}^z + p_{\bar{\nu}}^z)^2, \nonumber \\
    m_{t}^2 &=&  (E_b + E_{\ell^+} + E_\nu)^2 - (p_b^x + p_{\ell^+}^x + p_\nu^x)^2 - (p_b^y + p_{\ell^+}^y + p_\nu^y)^2 - (p_b^z + p_{\ell^+}^z + p_\nu^z)^2, \nonumber \\
    m_{\bar{t}}^2 &=& (E_{\bar{b}} + E_{\ell^-} + E_{\bar{\nu}})^2 - (p_{\bar{b}}^x + p_{\ell^-}^x + p_{\bar{\nu}}^x)^2 - (p_{\bar{b}}^y + p_{\ell^-}^y + p_{\bar{\nu}}^y)^2 - (p_{\bar{b}}^z + p_{\ell^-}^z + p_{\bar{\nu}}^z)^2, 
    \label{eq:conservation}
\end{eqnarray}
where $m_t = m_{\bar{t}} = 172.5$ GeV and $M_{W^+} = M_{W^-} = 80.4$ GeV are the pole masses of the top (anti)quark and $W^\pm$ boson respectively.  There are several methods to reconstruct the $t\bar{t}$ rest frame \cite{Barr:2010zj,Lester:1999tx,Cheng:2008hk,Cho:2008tj,Goncalves:2018agy}. In this analysis we use an analytical method  which aims at solving the quartic equation in neutrino momentum with the help of the Sonnenschein method \cite{Sonnenschein:2005ed}\footnote{We have implemented this method using Ref. \cite{Millar:2018dvh}. We have written the entire implementation that can be found in  this \href{https://github.com/declanmillar/x-tt-analysis}{github} repository in the \textsf{MadAnalysis}~5 framework.}. The number of unknowns in equation \eqref{eq:conservation} are the components of the neutrino momenta which are eight. First, we use equation \eqref{eq:conservation} to eliminate $E_\nu$ and $E_{\bar{\nu}}$. The longitudinal momenta $p_\nu^z$ and $p_{\bar{\nu}}^z$ can be further eliminated by other rearrangements of the two equations. Finally, we can enforce the conservation of the transverse missing momentum  to eliminate $p_\nu^y$ and $p_{\bar{\nu}}^y$, so we get
\begin{eqnarray}
    h_4 p_{x}^4 + h_3 p_x^3 + h_2 p_x^2 + h_1 p_x + h_0 = 0,
    \label{eq:quartic}
\end{eqnarray}
where $p_{x} = p_\nu^x$ and $h_0, h_1, h_2, h_3, h_4$ are numerical coefficients that depend on the four-momentum components of the bottom jets and charged leptons. The solution of this equation leads to a fourfold ambiguity. Removing the $b$-tagged jets that may results from QCD or the Higgs boson decay, we get two other possible combinations from matching the charged leptons with the $b$-jets and leading to an eightfold ambiguity. We can then obtain $p_\nu^z$ and $p_\nu^y$ from the arrangement we have used previously to eliminate these. The energy component can be then obtained from 
\begin{eqnarray}
    E_\nu &=& \sqrt{(p_\nu^x)^2 + (p_\nu^y)^2 + (p_\nu^z)^2}, \nonumber \\
    E_{\bar{\nu}} &=& \sqrt{(p_{\bar{\nu}}^x)^2 + (p_{\bar{\nu}}^y)^2 + (p_{\bar{\nu}}^z)^2}.
\end{eqnarray}
An important step is to choose the best suited solution for the neutrino momenta. There are different methods for making this choice: ({\it i}) solution that minimizes the invariant mass of the $t\bar{t}$ system; ({\it ii}) solution that uses the target mass of the $t\bar{t}$ resonance (which may be more suited for resonance searches);   ({\it iii}) characterization of the top quark decay products using kinematical information. We employ the latter method as it seems to be process-independent and have small biases. To do so we use the angular information encoded in the `lego' distance ($\Delta R$) between the (anti)top quark and its decay products and between the decay products themselves plus the ratios of the neutrino momenta with respect of the visible object momenta used in the reconstruction process. On a event-by-event basis, we construct these variables for each neutrino solution and we take the solution that maximizes the following Likelihood function:
\begin{eqnarray}
    {\cal L}_{\rm prob.} = \prod_i {P}(\Delta R(\nu, i)) \times { P}\bigg(\frac{p_T^\nu}{p_T^i}\bigg) \times {P}\bigg(\frac{E_\nu}{E^i}\bigg),
\end{eqnarray}
where $i$ refers to all charged leptons and $b$-tagged jets used in equation \eqref{eq:conservation} and ${P}$ is the probability assigned to each combination. We have found that our implementation yields very good results for the top (anti)quark invariant mass, both the individual ones and the one of the pair. We show the $m_t$ and $m_{t\bar{t}}$ resolution of our reconstruction in Fig. \ref{fig:tt:resolution}, where we compare the reconstructed masses against Monte Carlo (MC) truth information. We must stress out that the cumulative efficiency for top quark mass reconstruction is about $\approx 10\%$ for all the processes considered in this analysis.

%\clearpage
%%%%%%%%%%%%%%%%%%%%%%%%%%%%%%%%%%%%%%%%%%
\section{CP-sensitive observables}
\label{sec:observables}
%%%%%%%%%%%%%%%%%%%%%%%%%%%%%%%%%%%%%%%%%%

\begin{figure}[!t]
    \centering
    \includegraphics[width=0.32\linewidth]{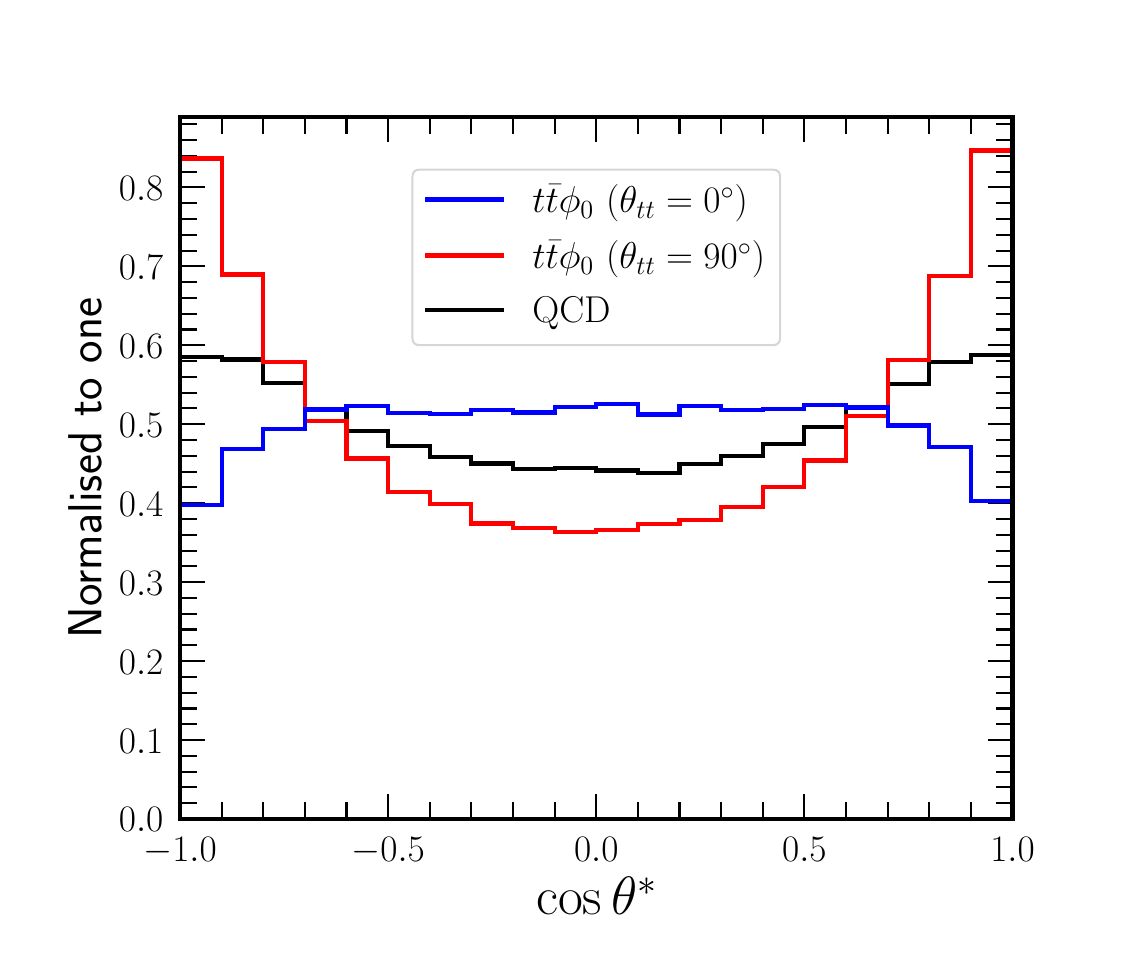}
    \includegraphics[width=0.32\linewidth]{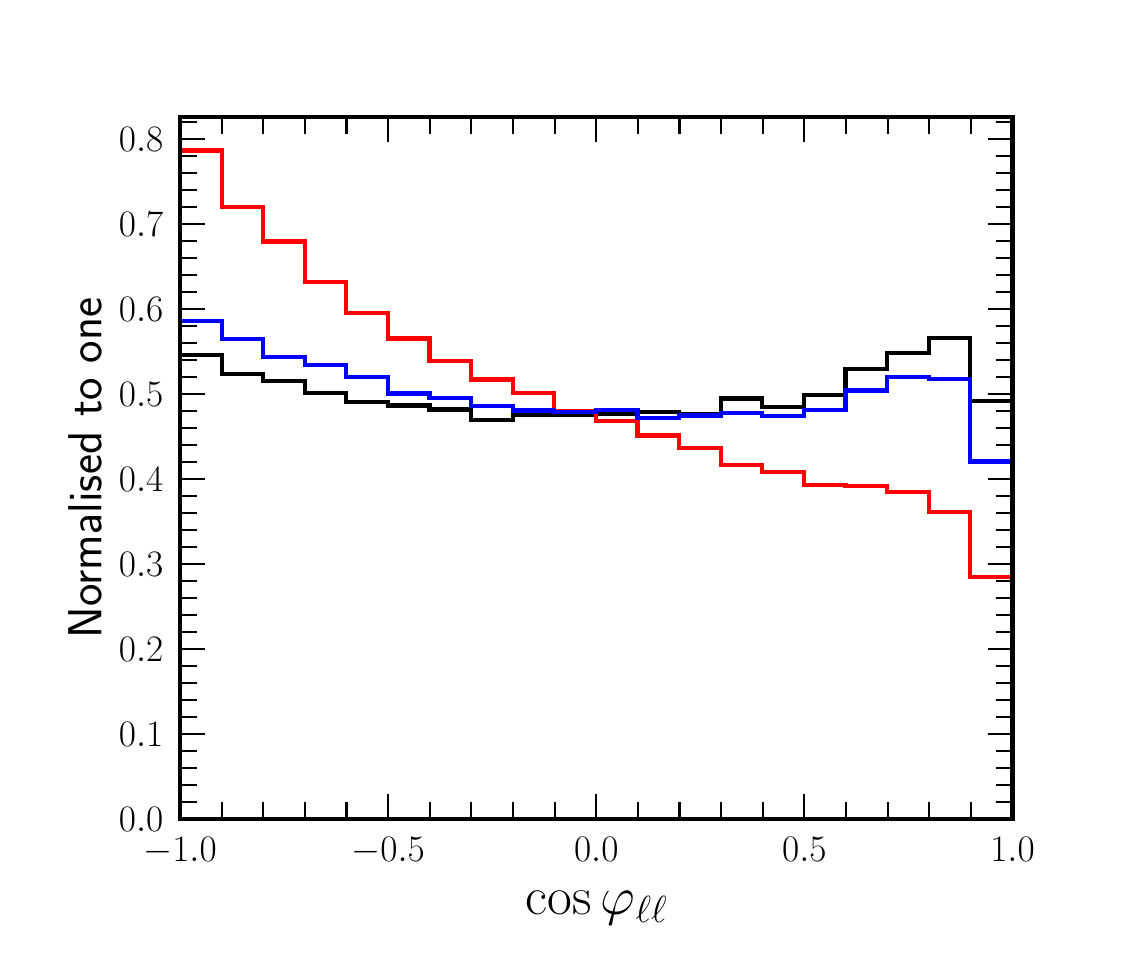}
    \includegraphics[width=0.32\linewidth]{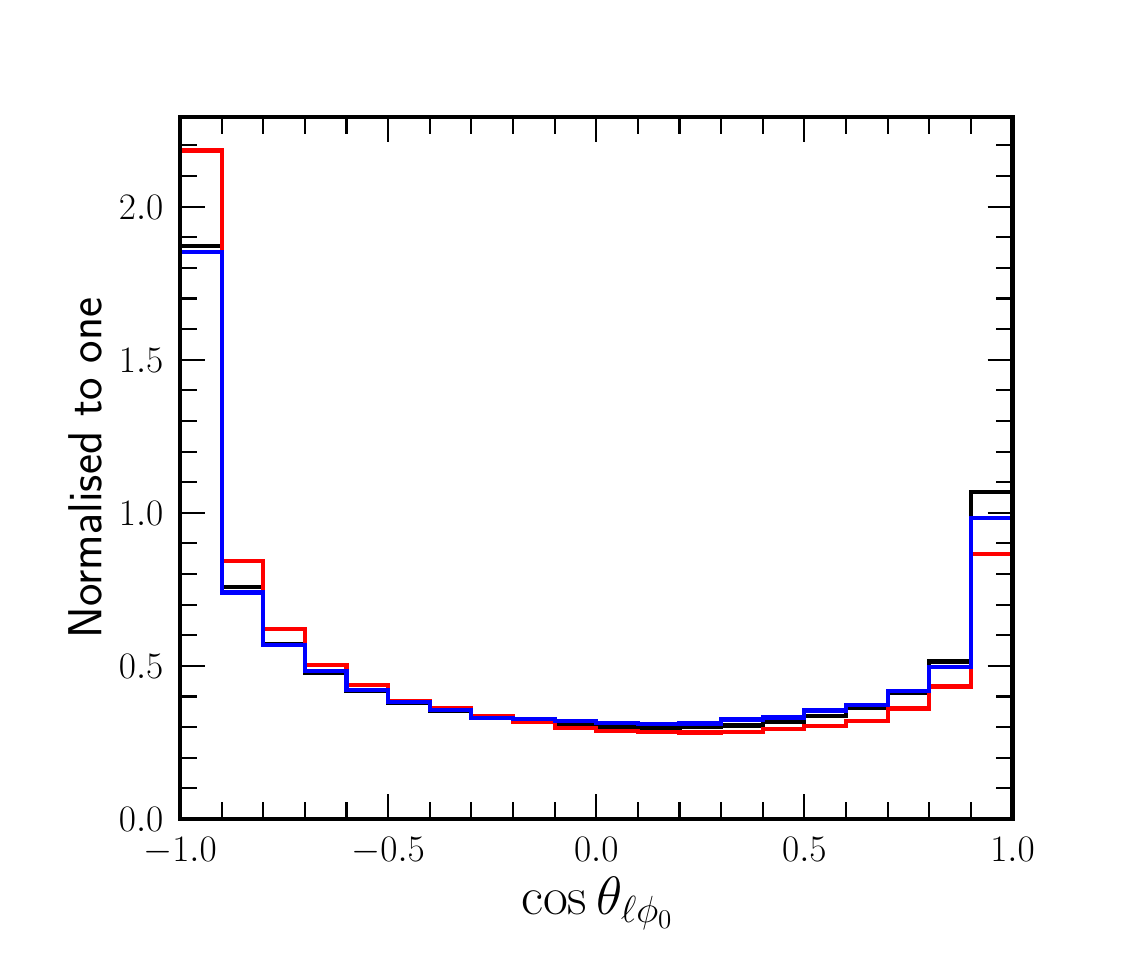}
    \vfill
    \includegraphics[width=0.32\linewidth]{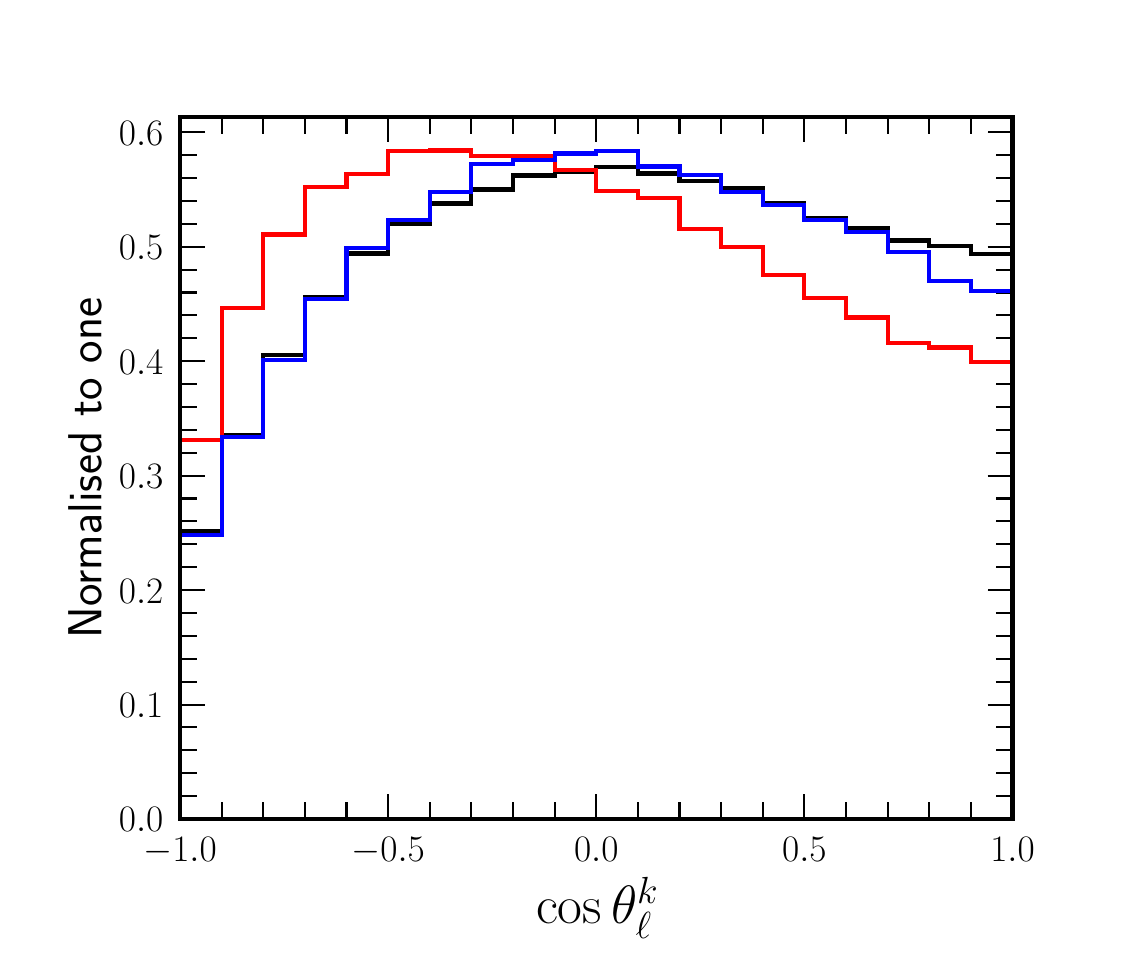}
    \includegraphics[width=0.32\linewidth]{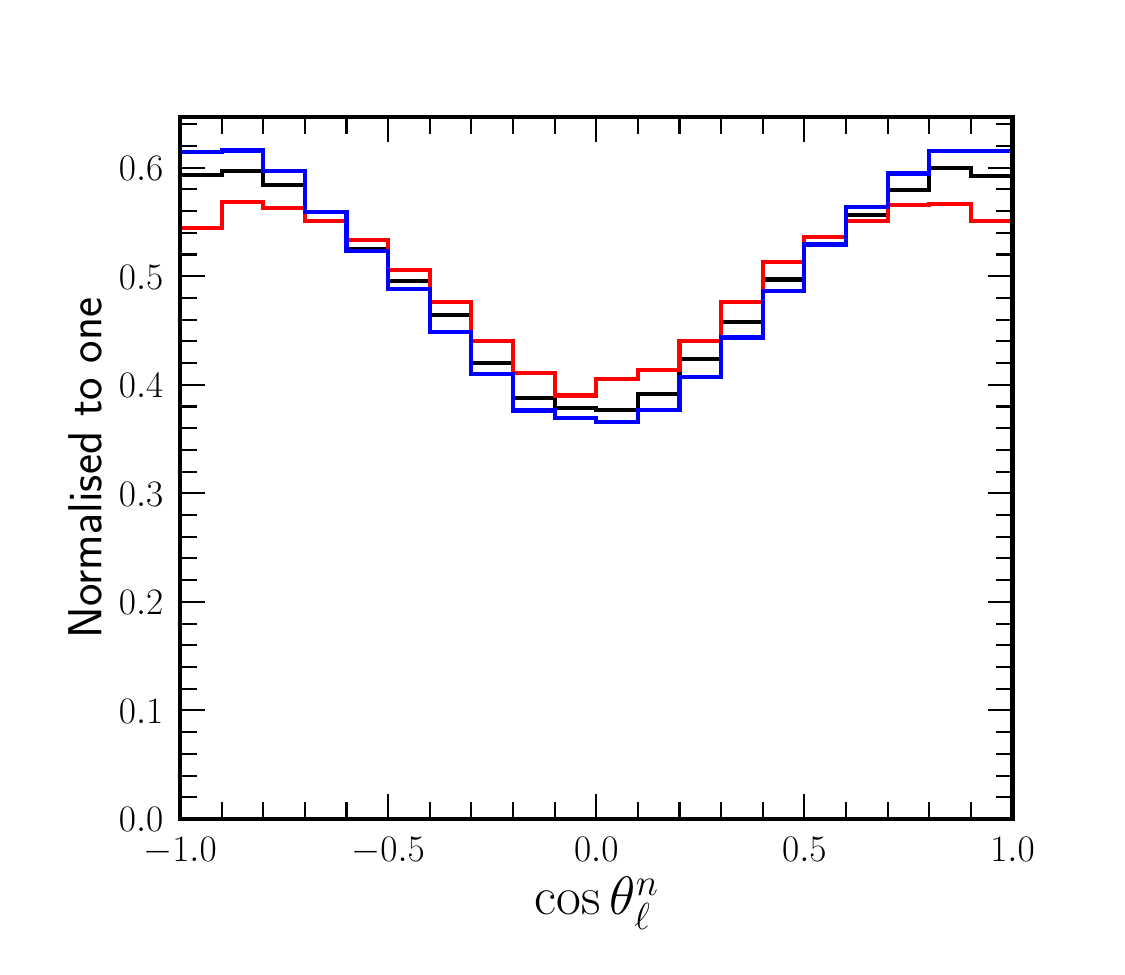}
    \includegraphics[width=0.32\linewidth]{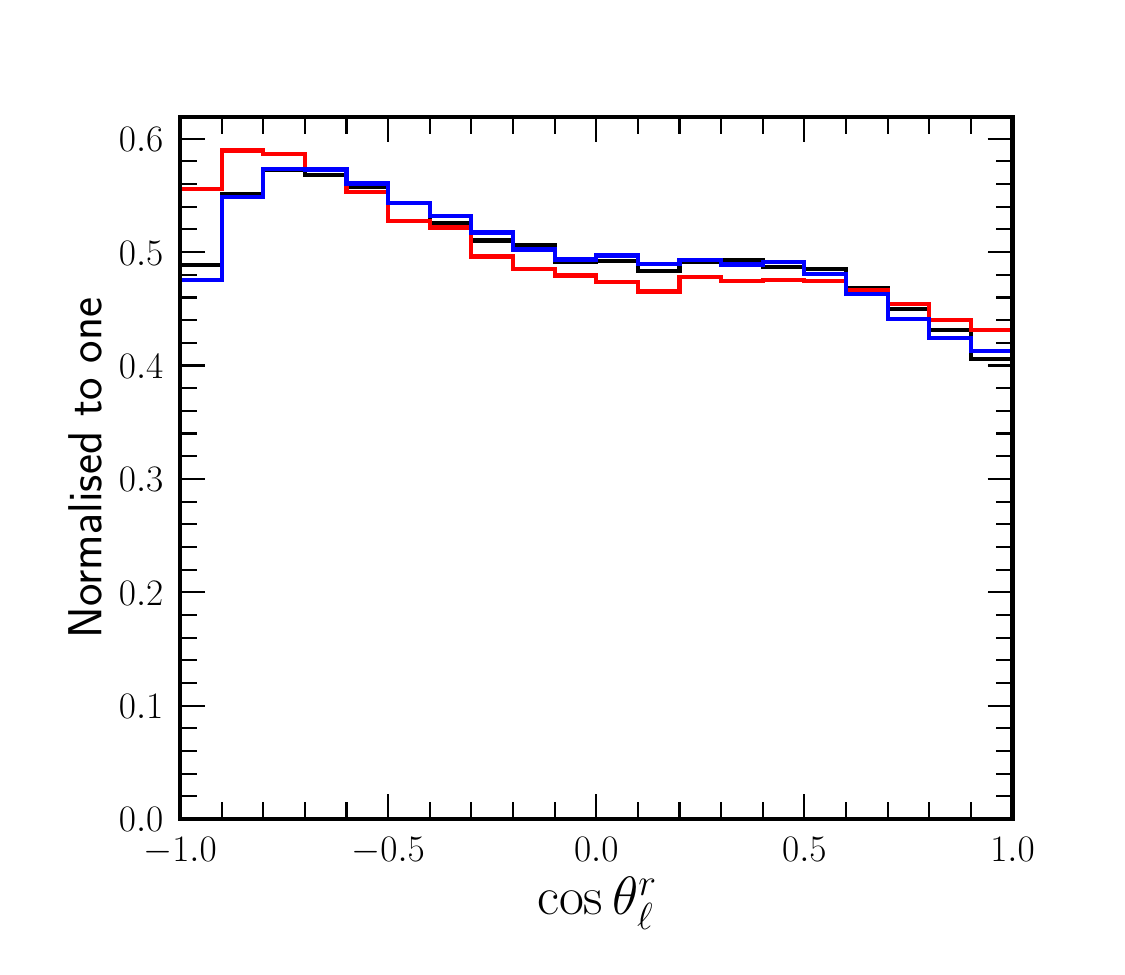}
    \vfill
    \includegraphics[width=0.32\linewidth]{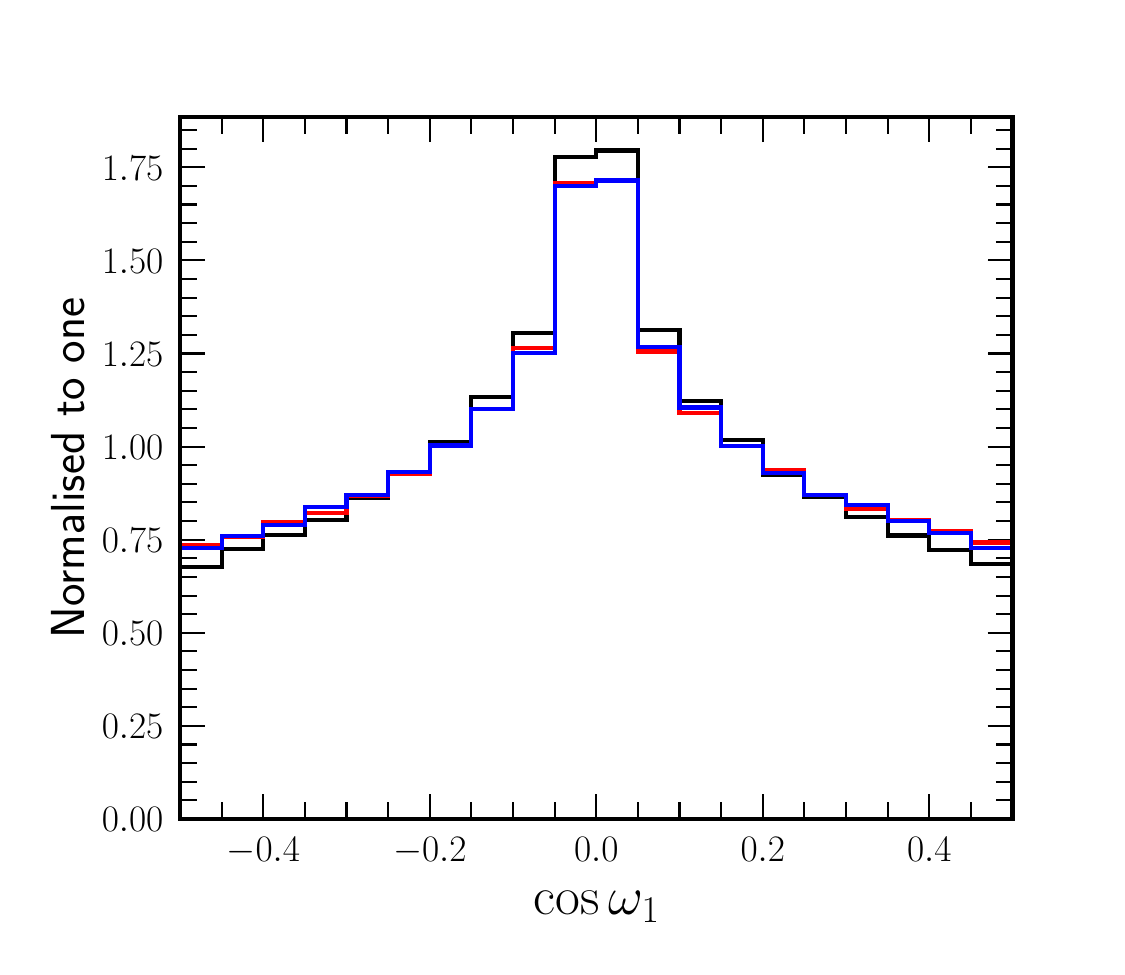}
    \includegraphics[width=0.32\linewidth]{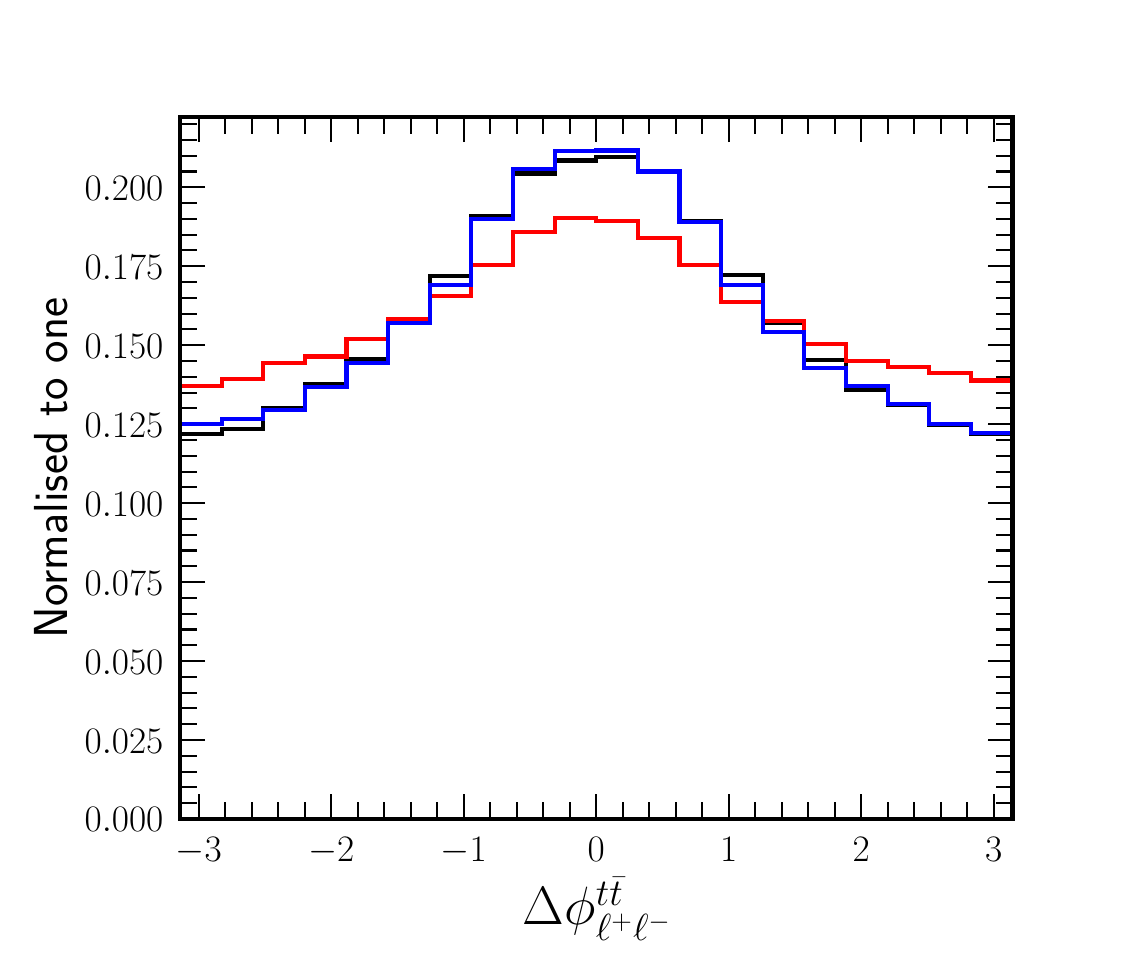}
    \includegraphics[width=0.32\linewidth]{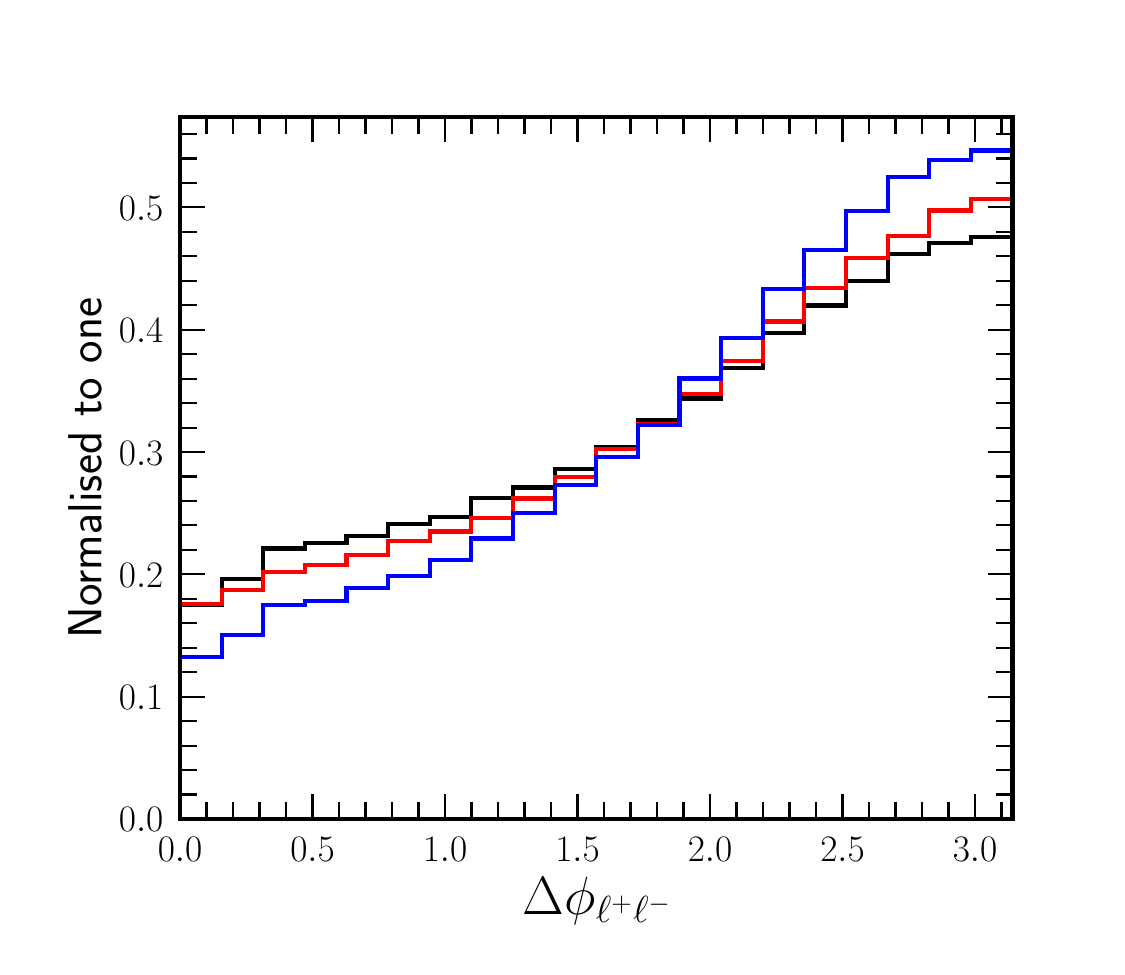}
    \caption{Examples of some distributions of the angles that are sensitive to the {\rm CP} nature of the $t\bar{t}\phi_0$ coupling for $t\bar{t}\phi_0$ with $\theta_t = 0\degree$ (blue), $t\bar{t}\phi_0$ with $\theta_t = 90\degree$ (red) and QCD backgrounds (black). In the upper panels we show $\cos\theta^*$ (left), $\cos\varphi_{\ell\ell}$ (middle) and $\cos\theta_{\ell\phi_0}$ (right). In the middle panels we show the cosine of the polar angle in the helicity basis (left), in the transverse basis (middle) and in the $r$--basis (right). The lower panels show $\cos\omega_1$ (left), $\Delta\phi_{\ell^+\ell^-}^{t\bar{t}}$ (middle) and $\Delta\phi_{\ell^+\ell^-}$ (right). All the distributions are normalised to unity and the calculations are done for $\kappa_t = 1$.}
    \label{fig:distributions}
\end{figure}

In this section, we briefly review the different angular and energy observables that we use in this study. In addition to low-level and high-level kinematics variables that we discuss in Sec. \ref{sec:kinematics} we also employ 39 angular and energy-ratio features (some distributions are shown in Fig. \ref{fig:distributions}). Note that these features have been studied extensively in the literature in different physical applications onto the top (anti)quark sector (see e.g. Refs. \cite{Mahlon:1995zn,Bernreuther:2013aga,Boudjema:2015nda,Shelton:2008nq,Godbole:2011vw, Rindani:2011pk,Prasath:2014mfa,Godbole:2015bda,Jueid:2018wnj,Arhrib:2018bxc, Godbole:2019erb,Arhrib:2019tkr,Chatterjee:2019brg,Goncalves:2018agy,Cheung:2020ugr, Faroughy:2019ird}).

\paragraph{Polarization and spin-spin correlation observables.} They consist of some polar angles of the charged leptons from (anti)top quark decays. The generic expression can be written as follows \cite{Mahlon:1995zn, Bernreuther:2013aga}:
\begin{equation}
   \frac{1}{\sigma} \frac{\text{d}^2\sigma}{\text{d}\cos\theta_{\ell^a} \text{d}\cos\theta_{\ell^b}} =
   \frac{1}{4}\bigg(1 + \alpha_{\ell^a} P_{a} \cos\theta_{\ell^a} + \alpha_{\ell^b} P_b \cos\theta_{\ell^b} + \alpha_{\ell^a} \alpha_{\ell^b} C_{ab} \cos\theta_{\ell^a} \cos\theta_{\ell^b}\bigg),
 \label{thetadouble}
\end{equation}
where $\alpha_\ell = \pm 1$ is the spin analysing power of the charged lepton and $\theta_{\ell^{a,b}} = \measuredangle (\hat{\ell}^{a,b}, \hat{S}_{a,b})$. Here, $\hat{\ell}^{a,b}$ refers to the direction of flight of the charged lepton in the top (anti)quark rest frame and $\hat{S}_{a,b}$ is the spin quantization axis in the basis $a$. We use three commonly studied bases\footnote{See, e.g., Refs. \cite{ATLAS:2016bac,CMS:2019nrx} for some recent measurements of spin-spin correlations in $t\bar{t}$ production.}.
\begin{itemize}
    \item Helicity basis ($a = k$): The spin quantization axis is defined as  the direction of motion of the top (anti)quark in the $t\bar{t}$ Zero-Momentum Frame (ZMF). In this case, $\cos\theta_{\ell+}^k$ is defined as
    \begin{eqnarray}
        \cos\theta_{\ell^+}^k = \frac{\hat{p}_{\ell^+} \cdot \hat{p}_t^{\rm ZMF}}{|\hat{p}_{\ell^+}||\hat{p}_t^{\rm ZMF}|}
    \end{eqnarray}
    and a similar expression  holds for $\cos\theta_{\ell^-}^k$.
    \item Transverse basis ($a = n$): In this case, the spin axis is defined as the three vector transverse to the production plane composed by the top quark direction of motion in the ZMF and the beam direction\footnote{As explained, in $t\bar{t}\phi_0$ production, the contribution of gluon fusion dominates the production rate which makes the initial state Bose symmetric. Therefore, and by following the recommendation of Ref. \cite{Bernreuther:2013aga}, the value of $\cos\theta_\ell^{n,r}$ is multiplied by the sign of the scattering angle $\vartheta = \hat{\textbf{p}}\cdot\hat{\textbf{p}}_t$ with $\hat{\textbf{p}}_t=\textbf{p}_t/|\textbf{p}_t|$  the top quark direction of flight in the $t\bar{t}$ ZMF and $\hat{\textbf{p}}=(0,0,1)$.}. Therefore, $\hat{S}_a$ is given by
    \begin{eqnarray}
        \hat{S}_a = \hat{p}_t^{\rm ZMF} \times \hat{p}_{\rm beam}.
    \end{eqnarray}
    \item $r$-basis ($a = r$): This basis is defined as the transverse to the plane defined by the top quark direction of flight in the ZMF and the spin axis in the transverse basis, such that the three vectors form a complete orthogonal basis. We also weight  $\cos\theta_{\ell^\pm}^r$ by the sign of the scattering angle between the (anti)quark and a unit vector $\hat{\bf p} = (0, 0, 1)$. 
\end{itemize}

The polarization of the (anti)top quark can be easily obtained by integrating \ref{thetadouble} over the angle $\theta_\ell^a$ (or $\theta_\ell^b$):
\begin{equation}
   \frac{1}{\sigma} \frac{\text{d}\sigma}{\text{d}\cos\theta_{\ell^\pm}^a} =
   \frac{1}{2}\bigg(1 + \alpha_{\ell^\pm} P_{t,\bar{t}}^a \cos\theta_{\ell^\pm}^a \bigg).
 \label{thetasingle}
\end{equation}

\paragraph{Laboratory frame observables.} We also consider some laboratory frame observables, {i.e.}, observables that are constructed from the particle momenta in the laboratory frame. First, the difference in the azimuthal angles of the two charged leptons is a clean observable that is usually used to measure spin-spin correlations between the top and the antitop quarks in $t\bar{t}$ production and decay. It was found in Ref. \cite{Cheung:2020ugr} that it can also serve as a good discriminator between the different {\rm CP} hypotheses of the $t\bar{t}\phi_0$ coupling. It is defined as 
\begin{eqnarray}
    \Delta \phi_{\ell^+ \ell^-}=|\phi_{\ell^+} - \phi_{\ell^-}|,
\end{eqnarray}
where $\phi_{\ell^\pm}$ is the azimuthal angle of the charged leptons in the laboratory frame. We also study the sensitivity of an observable that relies on the reconstruction of the Higgs boson candidate \cite{Boudjema:2015nda}:
\begin{eqnarray}
\cos\theta_{\ell \phi_0} = \frac{(\hat{p}_{\ell^+} \times \hat{p}_{\phi_0}) \cdot  (\hat{p}_{\ell^-} \times \hat{p}_{\phi_0})}{|(\hat{p}_{\ell^+} \times \hat{p}_{\phi_0})|  |(\hat{p}_{\ell^-} \times \hat{p}_{\phi_0})|},
\label{eq:costhetaHL}
\end{eqnarray}
with $\hat{p}_{\ell^+}$, $\hat{p}_{\ell^-}$ and $\hat{p}_{\phi_0}$ being  the directions of flight of the positively-charged, negatively-charged lepton and of the reconstructed Higgs boson candidate in the laboratory frame. Thus,  $\theta_{\ell \phi_0}$ defines the angle spanned by the dilepton system projected on the plane that is orthogonal to the Higgs boson candidate momentum. 

Another angle can be defined from $\theta_{\ell \phi_0}$, as follows: 
\begin{eqnarray}
\cos\tilde{\theta}_{\ell \phi_0} = \lambda \cos\theta_{\ell \phi_0},
\end{eqnarray}
with $\lambda = {\rm sign}((\hat{p}_{b} - \hat{p}_{\bar{b}})\cdot(\hat{p}_{\ell^-} \times \hat{p}_{\ell^+}))$ and $\hat{p}_b$ and $\hat{p}_{\bar{b}}$ the directions of flight of the $b$-(anti)quarks forming the Higgs boson candidate.

We also include some observables introduced in Ref. \cite{Faroughy:2019ird}, from where we show the definition of the most sensitive observable (and to where we refer the reader for more details), as follows: 
\begin{eqnarray}
    \cos\omega_1 &=& \frac{\hat{p}_{\phi_0}\cdot (\hat{p}_{\ell^+} \times \hat{p}_{\ell^-})~\hat{p}_{\phi_0}\cdot (\hat{p}_{\ell^+} - \hat{p}_{\ell^-})}{|\hat{p}_{\phi_0}|^2 |\hat{p}_{\ell^+} \times \hat{p}_{\ell^-}||\hat{p}_{\ell^+} -\hat{p}_{\ell^-}|}.
\end{eqnarray}

\paragraph{Energy-ratio observables.} Observables based on the ratios of the energies of the top (anti)quark and its decay products may carry some information on its polarization/helicity state (see, e.g., Refs. \cite{Shelton:2008nq,Godbole:2011vw, Rindani:2011pk, Prasath:2014mfa, Godbole:2015bda, Jueid:2018wnj, Arhrib:2018bxc, Godbole:2019erb, Arhrib:2019tkr,  Chatterjee:2019brg}). We define these as
\begin{equation}
u = \frac{E_\ell}{E_\ell+E_b}, \qquad
z = \frac{E_b}{E_t}, \qquad x_\ell = \frac{2 E_\ell}{m_t},
\end{equation}
with $E_\ell$, $E_b$ and $E_t$ being the energies of the charged lepton,
$b$-quark and top quark in the laboratory frame.

\paragraph{Other observables.} We start by considering the opening angle between the two oppositely charged leptons which is defined as 
\begin{eqnarray}
 \cos\varphi_{\ell_a\ell_b} = \frac{\hat{p}_{\ell^+} \cdot \hat{p}_{\ell^-}}{|\hat{p}_{\ell^+}| |\hat{p}_{\ell^-}|},
\end{eqnarray}
where $\hat{p}_{\ell^+}$($\hat{p}_{\ell^-}$) is the direction of flight of the charged lepton $\ell^+$($\ell^-$) in the $t\bar{t}$ ZMF. We also include two observables studied in Ref. \cite{Goncalves:2018agy}. The first observable is defined as the angle between the top quark direction of flight in the $t\bar{t}$ ZMF and the beam three-momentum, {i.e.},
\begin{eqnarray}
    \cos\theta^* = \frac{\hat{p}_{t}^{\rm ZMF} \cdot \hat{p}_{\rm beam}}{|\hat{p}_{t}^{\rm ZMF}| |\hat{p}_{\rm beam}|}.
    \label{eq:costhetaS}
\end{eqnarray}
The last observable, denoted by $\Delta\phi_{\ell^+ \ell^-}^{t\bar{t}}$, which is defined as the angle spanned by the direction of the two leptons on the orthogonal plane to the top quark 3-momentum in the $t\bar{t}$ ZMF
\begin{eqnarray}
    \Delta\phi_{\ell^+ \ell^-}^{t\bar{t}} = {\rm sign}[\hat{p}_t \cdot (\hat{p}_{\ell^+} \times \hat{p}_{\ell^-})] \arccos\bigg(\frac{(\hat{p}_{\ell^+}^{\rm ZMF} \times \hat{p}_{t}^{\rm ZMF}) \cdot (\hat{p}_{\ell^-}^{\rm ZMF} \times \hat{p}_{t}^{\rm ZMF})}{|\hat{p}_{\ell^+}^{\rm ZMF} \times \hat{p}_{t}^{\rm ZMF}| |\hat{p}_{\ell^-}^{\rm ZMF} \times \hat{p}_{t}^{\rm ZMF}|}\bigg).
\end{eqnarray}
%%%%%%%%%%%%%%%%%%%%%%%%%%%%
\section{Conditional DNNs}
\label{sec:DNNs}
%%%%%%%%%%%%%%%%%%%%%%%%%%%%
A conditional DNN for classification is a NN architecture where the classification process is conditioned on additional input information beyond the raw data. This additional information, often referred to as conditioning variables or features, can provide context or guidance to the NN, improving its ability to make accurate predictions. The network takes as input both the raw data to be classified and additional conditioning information. If the conditioning information is not directly compatible with the raw input data, it may need to be processed or transformed into a compatible format. This could involve feature extraction techniques such as encoding categorical variables, dimensionality reduction or any other preprocessing steps necessary to integrate the conditioning information with the input data. The network architecture is designed to incorporate the conditioning information into the classification process. This involves concatenating the conditioning information with the input data at later layers, passing it through additional conditioning layers to selectively focus on relevant parts of the input data based on the conditioning information, specifically, the angle $\theta_{tt}$. The network is trained to classify the input data into the appropriate  classes while taking into account the provided conditioning information. This enables the network to interpolate the results of different values of the conditional variable, $\theta_{tt}$, which the model did not trained on. In general, a conditional DNN is trained on a set of feature variables $x$ and a conditioned value $\theta_{tt}$, in which the network output is
\begin{eqnarray}
    \hat{y}  = \mathcal{F}(x,\theta_{tt})\,,
\end{eqnarray}
where $\mathcal{F}$ is the nonlinear function learned by the network to classify the input features $x$ conditioned by the value of $\theta_{tt}$. The training objective typically involves minimizing a classification loss function, such as cross-entropy loss, computed between the predicted class probabilities and the true class labels, while also considering any regularization terms to prevent overfitting.

\begin{table}[!thb]
    \setlength\tabcolsep{0.6 cm}
    \centering
    \begin{tabular}{lcc}
        \toprule
          & MLP  & GNN \\
        \midrule
        Number of hidden layers & 3 FC & 3 GNN + 6 FC \\
        Input layer dimension & 103 neurons & 11 nodes-fully connected    \\
        Output layer dimension & 1 neuron & 1 neuron \\
        Output layer activation & sigmoid &sigmoid \\
        Drop out rate & $20\%$ & $25\%$ \\
        Loss function & BCE& BCE \\
        Optimizer & Adam  & Adam \\
        Initial learning rate & 0.0001 & 0.0001 \\
        Epochs & 20 & 20  \\
        Batch size & 500 & 500  \\
        \bottomrule
    \end{tabular}
    \caption{\label{tab:structure}%
        Hyperparameters of the different DL networks. FC refers to Fully Connected layers while BCE denotes Binary Cross-Entropy loss.
         }\label{tab:hyp}
\end{table}
 In this paper, we utilize two conditional DNNs with different structure, namely, a conditional MLP and a  conditional (multi-modal) Graph Neural Network (GNN).The hyperparameters of both networks are summarized in Table  \ref{tab:hyp}. Both networks are trained on four signal points with $\theta_{tt}=0^\circ,\pm 30^\circ,\pm 45^\circ,\pm 90^\circ$ and interpolate the results for $\theta_{tt}=\pm 15^\circ,\pm 60^\circ,\pm 75^\circ$\footnote{Note that the conditional DNNs used in this paper are not sensitive to the sign of $\theta_{tt}$. This feature was also checked by comparing the angular distributions and total rates for different values of $\theta_{tt}$ across the region of consideration: $\theta_{tt}~\in~[-\pi/2, \pi/2]$}.

%%%%%%%%%%%%%%%%%%%%%%%%%%%%%%%%%%%%%
\subsection{MLP}
%%%%%%%%%%%%%%%%%%%%%%%%%%%%%%%%%%%%%%
Commencing with high-level kinematical distributions, we utilize a MLP model to enhance the discrimination between signal and background distributions. These distributions encapsulate distinctive information regarding the overall structure of both signal and background events. Consequently, the architecture of the MLP network, comprising fully connected layers, can discern global features effectively, resulting in robust classification power between signal and background events. Moreover,  the fully connected layers that process the kinematical features are comprised by one linear layer which encodes the values of the condition parameter $\theta_{tt}$. In this case, the MLP is able to learn  global features of the signal and background events that can be used to increase the signal to background yield assigned to each value of $\theta_{tt}$, which then enables the model to interpolates the values of $\theta_{tt}$ that the network is not trained on it. Accordingly, the MLP is trained on specific values of the angle and is used to provide the sensitivity to all other values of $\theta_{tt}$.

Despite the MLP capability to achieve high classification performance, the similarity in kinematical structures between certain background distributions and signal ones diminishes the overall classification efficiency. However, by implementing initial cuts that maximize the signal-to-background yield prior to inputting the distributions into the MLP, one can augment the classification performance. However, the constructed kinematical spectra demonstrate significant intercorrelation, such that applying a cut to any distribution may influence the structure of others, consequently impeding the MLP classification performance. To mitigate the global impact of these initial cuts, it is imperative to decorrelate such dependencies across kinematical variables, either via the square-root of the covariance matrix or Gaussian transformation of variables, as elucidated in \cite{TMVA:2007ngy}. Ultimately, while initial cuts may bolster classification performance, we have chosen not to apply these, thereby affording the MLP with complete autonomy in identifying optimal classification boundaries.

The MLP structure consists of two input layers. The first input layer is used to encode the features with $103$ neurons, which is the number of the used features. The second input layer encodes the value of the condition parameter $\theta_{tt}$ and it has only one neuron. The first input layer is followed, sequentially, by   three fully connected layers with $256,~128$ and $64$ neurons and ReLU activation function. Each fully connected layer is followed by a dropout layer with dropout rate of $20\%$. The second input layer is followed by one linear layer\footnote{It is important to keep the mapped conditioned parameter without any activation function.} with $64$ neurons, to adjust the dimension with the last fully connected layer from the first stream. The final layers from the two streams are concatenated using a concatenation layer where the output is passed directly to an output layer with one neuron and sigmoid activation function. For the optimization process, we employed a learning rate of \(1 \times 10^{-4}\) and a weight decay parameter of \(1 \times 10^{-4}\) for Adam optimizer to minimize a binary cross entropy loss function. 
%%%%%%%%%%%%%%%%%%%%%%%%%%%%%%%%%%%%%%%%%%%%%%%
\subsection{GNN}
%%%%%%%%%%%%%%%%%%%%%%%%%%%%%%%%%%%%%%%%%%%%%%%%
GNNs represent a class of DL models specifically designed for processing graph-structured data, where a graph is a set of nodes/vertices connected by edges $G(V,E)$. By effectively leveraging the inherent connectivity and relational information within graphs, GNNs excel in capturing complex patterns that are not easily accessible to traditional NN architectures. Central to their operation is the message passing mechanism, where node representations are iteratively updated by aggregating features from their neighboring nodes, thus encoding both local and global graph structures into the learning process. The versatility of GNNs is evident in their wide range of applications across various tasks. In graph classification, GNNs aim to predict the labels of entire graphs based on their structure and node features, which is crucial in domains like chemical compound analysis and social network studies. Node classification, another prominent task, involves predicting the labels of individual nodes within a graph, often used in citation networks and social media to identify categories or communities. Additionally, GNNs are employed in link prediction to forecast potential connections between nodes, which has significant implications for recommender systems and network analysis. Through these tasks, GNNs demonstrate their powerful capability to model complex relational data effectively. 
GNNs can be broadly categorized into several types based on their architecture and the specific methods they employ for node information aggregation. For instance, Graph Convolutional Networks (GCNs) \cite{kipf2017semisupervised} utilize a convolutional approach, adapting the traditional convolution operations to work directly on graphs. GCNs can extract meaningful features from nodes and edges, achieving state-of-the-art performance in graph classification tasks. This adaptability and capability to model complex relationships make GCNs a versatile and promising approach in HEP analysis.

\begin{figure}[!t]
    \centering
    \includegraphics[scale=0.7]{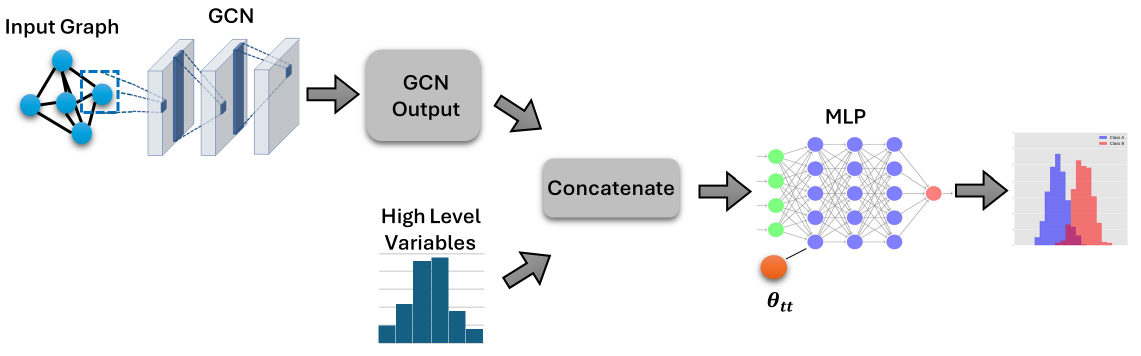}
    \caption{Schematic diagram of the (multi-modal) GNN, illustrating the flow from the input graph through the GCN to generate the final output. This output is concatenated with the high-level variables and processed through a MLP, this resulting in the final output classification.}
    \label{fig:gnn_architecture}
\end{figure}

For the purpose of constructing a conditional GNN, our approach integrates a multi-modal  network combining GCN with MLP. The GCN component is particularly advantageous for incorporating the topological relationships among nodes and edges, thereby facilitating the learning of graph-structured data. In our methodology, we represent the reconstructed particles, their daughters, and parent particles as nodes within the graph. Following the methodology outlined in \cite{Esmail:2023axd}, each node $i$ in the input layer is represented as a feature vector $x = (I_1, I_2, I_3, I_4, p_T, E, \eta, \phi)$. This vector encapsulates the properties of the corresponding particle, where $p_T$ denotes transverse momentum, $E$ is the energy, $\eta$ represents pseudorapidity and $\phi$ is the azimuthal angle. Initially, the values of $I_1$ through $I_4$ are set to zero. The indicator $I_1$ is set to 1 if the particle is a lepton, $I_2$ is set to 1 if the particle is a $b$-jet, $I_3$ is set to 1 if it is a neutrino and $I_4$ is set to 1 if it is a reconstructed top (anti)quark (and 0 otherwise,  in all cases). The input graph to the GCN consists of 11 nodes, specifically 2 leptons, 4 $b$-jets, 2 neutrinos, $2$ top (anti)quarks and the Higgs boson, each characterized by the aforementioned feature vector $x$. The graph is fully connected, with edges weighted by the angular distance $\Delta R_{(x_{i}, x_{j})}$ between the particles in nodes $i$ and $j$, thereby enabling the model to capture intricate spatial relationships between such particles.
Our model consists of 3 GCN layers, as first introduced in \cite{kipf2017semisupervised}, with ReLU activation. This is followed by max pooling to aggregate the node embedding. The other part of the model is a basic set of 3 fully connected layers (another MLP), which takes the high-level variables mentioned earlier that are used to train the basic MLP model. In addition, it also takes the output of the GCN model. Both inputs are concatenated and fed into the 3 fully connected layers. Our hybrid approach is trained conditionally similar to the basic MLP on specific values of the angle and is used to provide the sensitivity to all other values of \(\theta_{tt}\). The architecture schematic diagram, depicting all inputs and outputs, is shown in Fig. \ref{fig:gnn_architecture}.

For the optimization process, we employed a learning rate of \(1 \times 10^{-4}\) and a weight decay parameter  of \(1 \times 10^{-4}\) for Adam optimizer to minimize a binary cross entropy loss function. These values were selected to balance efficient learning with the stability of the model.  All models developed in our study were constructed using the \textsc{PyTorch Geometric} framework \cite{fey2019fast}, a powerful and efficient library designed to facilitate the implementation of graph-based DL models.

%%%%%%%%%%%%%%%%%%%%%%%%%%%%%%%%%%%%%%%%%%%%%%%%%%
\section{DL analysis}
\label{sec:DL}
%%%%%%%%%%%%%%%%%%%%%%%%%%%%%%%%%%%%%%%%%%%%%%%%%%
In this section we discuss the analysis taking place through the MLP and GNN models described. Starting with low level information about the reconstructed final state particles, we discuss how all features for the two DNNs are considered. 
%%%%%%%%%%%%%%%%%%%%%%%%%%%%%%%%%%%%%%%%%%%
\subsection{Signal and background kinematics}%
\label{sec:kinematics}
%%%%%%%%%%%%%%%%%%%%%%%%%%%%%%%%%%%%%%%%%%%

Considering the leptonic decays of the top (anti)quark and those of the Higgs boson into $b\bar{b}$, the final state will consist of two charged leptons ($\ell=e,\mu$) with opposite electric charge, at least four $b$-tagged jets and missing energy. The main background processes arise from the QCD production of $t\bar{t}$ in association with two jets ($t\bar tb\bar{b}$, $t\bar tjj$). As our analysis relies on exactly four $b$-tagged jets, background processes such as $t\bar{t}$, multi-jets and $V$+jets are subleading. We first apply some basic generator-level cuts on parton-level objects like electrons, muons, partons and missing energy, {i.e.}:
\begin{eqnarray}
    p_T(\ell) > 20~{\rm GeV}, ~|\eta_\ell| < 2.5, ~p_T(j) > 25~{\rm GeV}, |\eta_j| < 2.5,~~ E_{T}^{\rm miss} > 20~{\rm GeV}.
\end{eqnarray}

At the reconstruction level, we require that events do not contain any isolated hadronically-decaying $\tau$ lepton with $p_T > 30$ GeV and $|\eta| < 2.5$. Then we require exactly two charged leptons (electrons or muons) with $p_T > 25$ GeV and $|\eta| < 2.4$ excluding electrons in the transition region in the calorimeter ({i.e.}, those with $1.44  < |\eta| < 1.57$). The charged leptons are required to be isolated using tight isolation criteria. To achieve this we have calculated a variable ($I_{\ell}$) defined as 
\begin{eqnarray}
I_{\ell} = \sum_{i \in \Delta R} p_{T,i},
\end{eqnarray}
with the sum being over all the tracks within $\Delta R = 0.3$ of the charged lepton excluding the lepton candidate itself. A signal lepton is defined to satisfy $I_\ell < 10$ GeV. Furthermore, we require overlap removal between different objects in the events, {\it i.e.} we require that jets and leptons to be separated by $R > 0.4$. We then require that $E_{T}^{\rm miss} > 30$ GeV. We impose that events should contain at least four $b$-tagged jets with $p_T > 30$ GeV and $|\eta| < 2.4$. The combination of the two $b$-tagged jets in the event that match the charged leptons will be used for the top quark reconstruction as described in Sec. \ref{sec:reconstruction} while the remaining $b$-tagged jets are ordered in transverse momentum for which case the two leading $b$-tagged jets are used to form $\phi_0$ candidates. We require that the $t\bar t$ invariant mass to lie in the window of $[132.5, 210.5]$ GeV. We, however, do not impose any requirement on the invariant mass of the $\phi_0$ candidate. After this basic selection, the signal significance for $\theta_{tt}=0^\circ$ and $\kappa_{tt}=1$ is $0.5$~($1.7$) assuming $300$~($3000$) fb$^{-1}$ of luminosity.  

After all the events pass the basic selection criteria, we explore the following variables for more sophisticated DL analyses.

\paragraph{${\rm CP}$-sensitive variables.} They consist of 39 variables which were described in detail in Sec. \ref{sec:observables}. These observables can be split into two categories: ({\it i}) ${\rm CP}$-even observables which do not change with $\theta_{tt}$ and ({ii}) ${\rm CP}$-odd observables that are sensitive to the change in $\theta_{tt}$. The classification of the angular observables is shown in Table \ref{tab:CP:classification}.

\begin{table}[!t]
\setlength\tabcolsep{8pt}
    \centering
    \begin{tabular}{l c}
    \toprule
    	${\rm CP}$-even observables & $\cos\theta_{\ell}^k \cos\theta_{\ell}^k$, $\cos\theta_{\ell}^n \cos\theta_{\ell}^n$, $\cos\theta_{\ell}^r \cos\theta_{\ell}^r$ \\ [0.4em]
	& $\cos\theta_{\ell}^r \cos\theta_{\ell}^n$, $\cos\theta_{\ell}^r \cos\theta_{\ell}^k$, $\cos\theta_{\ell}^k \cos\theta_{\ell}^n$, $x_\ell$, $u$, $z$  \\ [0.4em]
	& $\cos\omega_1$, $\cos\omega_2$, $\cos\omega_3$, $\cos\omega_4$, $\cos\omega_5$, $\cos\omega_6$ \\ 
    $\cos\theta_{\ell}^k$, $\cos\theta_\ell^n$, $\cos\theta_\ell^r$, $\cos\varphi_{\ell\ell}$, $\Delta\phi_{\ell\ell}$, $\Delta\phi_{\ell\ell}^{t\bar{t}}$ \\
	\toprule
	${\rm CP}$-odd observables & $\cos\theta^*$, $\cos\theta_{\ell\phi_0}$, $\cos\tilde{\theta}_{\ell\phi_0}$ \\  [0.4em]
       \bottomrule
    \end{tabular}
    \caption{Classification of angular observables that we use in this analysis. }
    \label{tab:CP:classification}
\end{table}

\paragraph{Low-level variables.} Low-level variables consist of the four components of the momenta of the top (anti)quarks and Higgs boson decay products. There are 32 variables in total, as follows.
\begin{itemize}
    \item The two neutrino candidates from the solution of the Likelihood-based top (anti)quark reconstruction method (see Sec. \ref{sec:reconstruction}): 
    $$
    \eta(\nu_1), p_T(\nu_1), E(\nu_1), \phi(\nu_1),\eta(\nu_2), p_T(\nu_2), E(\nu_2), \phi(\nu_2).
    $$
    \item The four momenta of the two charged leptons: 
    $$
    \eta(\ell_1), p_T(\ell_1), E(\ell_1), \phi(\ell_1),\eta(\ell_2), p_T(\ell_2), E(\ell_2), \phi(\ell_2).
    $$
    \item The four momenta of the four $b$-tagged jets: 
    \begin{eqnarray*}
    \eta(b_1), p_T(b_1), E(b_1), \phi(b_1),\eta(b_2), p_T(b_2), E(b_2), \phi(b_2), \\
    \eta(b_3), p_T(b_3), E(b_3), \phi(b_3),\eta(b_4), p_T(b_4), E(b_4), \phi(b_4).
    \end{eqnarray*}
\end{itemize}

\paragraph{High-level variables.} They consist of more complicated variables, which are built upon the low-level variables. There are 30 variables in total, as follows. 
\begin{itemize}
    \item Invariant mass of the top quark,  top antiquark and Higgs boson candidates: $m_{t_1}, m_{t_2}$ and $m(b_3 b_4)$.
    \item The components of the four momenta of the top quark, top antiquark and Higgs boson candidates:
    \begin{eqnarray*}
        \eta(t_1), p_T(t_1), E(t_1), \phi(t_1), \\
        \eta(t_2), p_T(t_2), E(t_2), \phi(t_2), \\
        \eta(b_3b_4), p_T(b_3 b_4), E(b_3 b_4), \phi(b_3 b_4).
    \end{eqnarray*}
    \item The invariant mass, the energy and the transverse momentum of the $t\bar{t}$ system:
    \begin{eqnarray*}
        m(t_1 t_2), p_T(t_1 t_2), E(t_1 t_2).
    \end{eqnarray*}
    \item The invariant mass of the $t\bar{t}\phi_0$ system: $m(t_1t_2b_3b_4)$.
    \item The transverse of the dilepton system: $p_T(\ell_1\ell_2)$.
    \item The scalar sum of the jet transverse momenta including all the $b$-tagged jets in the events
    \begin{eqnarray*}
        H_T^b \equiv \sum_i p_T(b_i).
    \end{eqnarray*}
    \item The effective mass
    \begin{eqnarray*}
        M_{\rm eff} \equiv H_T^b + p_T(\ell_1) + p_T(\ell_2).
    \end{eqnarray*}
    \item The minimum and maximum of the transverse momentum and the invariant of the top (anti)quark and the $b$-tagged jets forming the Higgs boson candidates:
    \begin{eqnarray*}
        p_T^1 \equiv \max_i\{p_T(b_i, t_1), p_T(b_i, t_2)\}, \quad 
        p_T^2 \equiv \min_i\{p_T(b_i, t_1), p_T(b_i, t_2)\}.
    \end{eqnarray*}
\end{itemize}

%%%%%%%%%%%%%%%%%%%%%%%%%%%%%%%%%%%%%%%%%%%%%%%%%%%%%%%%
\subsection{MC event generation and simulation tools}%%
%%%%%%%%%%%%%%%%%%%%%%%%%%%%%%%%%%%%%%%%%%%%%%%%%%%%%%%%
The effective Lagrangian of Sec. \ref{sec:model} is implemented in \textsf{FeynRules} \cite{Alloul:2013bka}. The output file in UFO format \cite{Degrande:2011ua} is used as an input to \textsf{MadGraph5\_aMC@NLO} \cite{Alwall:2014hca} to generate parton-level samples for both the signal and  background events. As mentioned, all  signal and background processes are simulated at LO in QCD. To keep track of spin and correlation effects, we decay the intermediate resonances by using \textsf{MadSpin} \cite{Artoisenet:2012st}. \textsf{Pythia}~version 8309 is used to add parton showering, hadronization and heavy hadron decays to the event samples \cite{Bierlich:2022pfr}. The detector response is modelled using the Simplified Fast-detector Simulator (SFS) \cite{Araz:2020lnp} in the \textsf{MadAnalysis}~5 framework \cite{Conte:2012fm,Conte:2014zja,Conte:2018vmg,Araz:2019otb}. The simulation of the reconstructed objects such as tracks, isolated electrons, jets, hadronically-decaying $\tau$ lepton and missing transverse energy ($E_{T}^{\rm miss}$) is done following the same lines of Ref. \cite{Frank:2023epx}, which is based on a CMS analysis targeting the search of right-handed gauge bosons in the two leptons and two jet events \cite{CMS:2021dzb}. We have, however, slightly modified the detector card in this analysis by assuming a flat $85\%$ $b$-tagging and $10\%$ mis-tagging efficiencies across all the pseudorapidity and transverse momentum values. Jets are clustered with the anti-$k_t$ algorithm \cite{Cacciari:2008gp} with a jet radius of $R=0.4$ using \textsf{FastJet} version 3.4.1 \cite{Cacciari:2011ma}. 

As mentioned, for the DNN analysis, we use \textsf{PyTorch Geometric} \cite{fey2019fast} for building the GNN network, while standard \textsf{PyTorch} \cite{paszke2019pytorch} is used for the MLP. Finally, the  \textsf{Scikit-Learn} package \cite{pedregosa2011scikit} is used to facilitate  network training and evaluation. 

%%%%%%%%%%%%%%%%%%%%%%%%%%%%%%%%%%%%%
\subsection{Training the network}%%%%%%%%%%
%%%%%%%%%%%%%%%%%%%%%%%%%%%%%%%%%%%%%
Once the datasets are prepared, we commence training the networks to understand the complex, non-linear relationships between the input data and their corresponding labels. Additionally, the network learns to interpolate between the trained values of the conditioned parameter. Events are organized into a feature dataset with dimensions $(n,103)$, where $n$ represents the number of events in the training dataset, and a conditional input vector with the value of $\theta_{tt}$.
The training dataset is composed of equally-sized events generated with $\theta_{tt} = 0^\circ,\pm 30^\circ,\pm 45^\circ$, and $\pm 90^\circ$, each containing $100000$ events, resulting in a training signal dataset of size $400000$ events. The conditional parameter vector is prepared with the exact value of $\theta_{tt}$ in radians to facilitate better convergence of the network to the minimum of the loss function.

For the background dataset, we merge events from all background channels into a single dataset, weighting them according to their respective cross-sections and selection efficiencies. The conditional parameter input is a vector of the same length as the training background dataset, containing random values between $0$ and $\pi/2$. Since the primary objective of the network is to learn a global pattern of the signal events and interpolate between the trained values, we utilize training datasets of size $400000$ and $300000$ for both signal and background, respectively. This strategy is akin to assigning higher weights to the signal events, enabling the network to focus more on learning the features of the signal events.

For network evaluation, we utilize equal-sized, new, unseen datasets for the signal and background, each consisting of $100000$ events.
For each of the  training datasets we assign the label $Y=1$ for signal events while for background events we assign the label $Y=0$. In order to remove the network dependence on the position of the signal and background events,  we stack the signal and background events in one data set and shuffle it together with the assigned labels. During the network training stage, during each epoch (defined as number of passes of the entire data sets), the network updates the weights assigned to the neurons for each event via backward propagation of errors.  The network then tries to minimize the error between its predictions and the true labels by reaching a global minimum of some loss function. For this purpose we use a binary cross-entropy as a loss function and a Adam optimizer to optimize the network convergence to the global minimum of the loss function. Finally, the network repeats the process until it reaches the desired accuracy. Once the model is trained, we test it by using completely unseen new data sets to measure the network performance. 

We prepared seven test datasets according to the value of $\theta_{tt} = 0^\circ,\pm 15^\circ,\pm 30^\circ,\pm 45^\circ,\pm 60^\circ,$ 
$\pm 75^\circ$ and  $\pm 90^\circ$. These datasets are prepared of equal size of the signal events and background events. We stress here, that the network is trained on signal events with $\theta_{tt}= 0^\circ,\pm 30^\circ,\pm 45^\circ$ and $\pm 90^\circ$, and it used to interpolate the points in between. 

For all networks, we train the model with 20 epochs with batch size equalling a  $500$ sample. The dimension of the final output probability, $\hat{Y}$,  is $1\times 2$, $({P}_{sig},{P}_{bkg})$, with ${P}$ ranging between $[0,1]$. If ${P}_{sig} > 0.5\ ({P}_{bkg} < 0.5)$, the corresponding event is classified as most likely being a signal event and if ${P}_{sig} < 0.5 \ ({P}_{bkg} > 0.5)$ the corresponding event is classified as most likely being a background event.

\begin{figure}[!th]
    \centering
    \includegraphics[width=\textwidth]{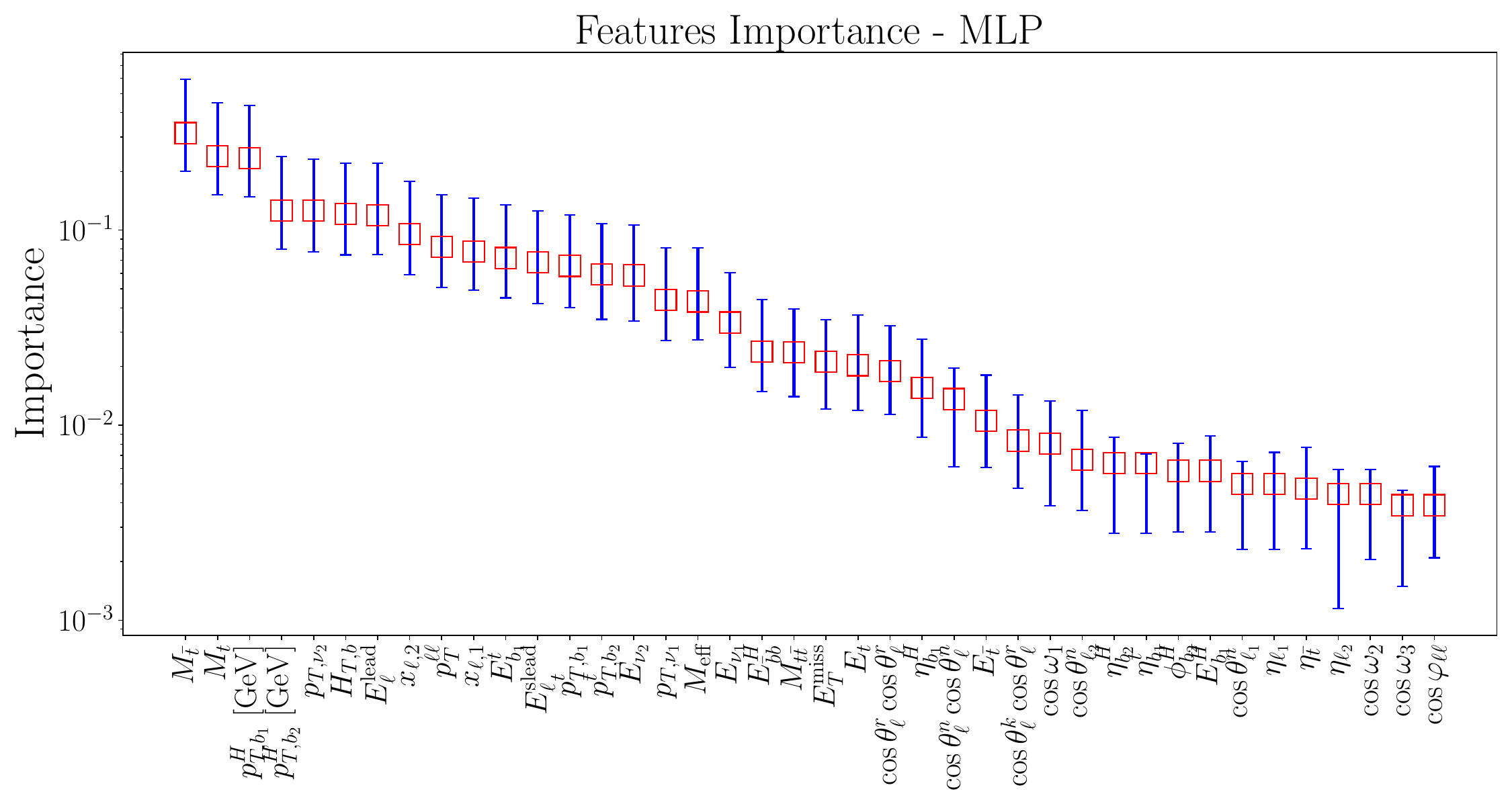}
    \caption{Feature importance for the top 40 ranked input variables in the MLP.  Y-axis shows the performance reduction when each variable is removed from the input dataset. The blue bars indicate the  maximum and minimum values obtained by replacing each variable with random numbers over five iterations, while the red boxes represent the median value.}
    \label{fig:f_importance}
\end{figure}
Finally, to check the importance of each of the input feature we compute the reduction of the MLP testing accuracy when the feature is replaced by random numbers. Figure \ref{fig:f_importance} represents the features importance for the 40 highly ranked variables  when each input variable is replaced by random numbers.  
%%%%%%%%%%%%%%%%%%%%%%%
\section{Results}
\label{sec:results}
%%%%%%%%%%%%%%%%%%%%%%%
In this section we present the results of using a conditional MLP and GNN to probe the CPV phase in $pp\to t\bar{t}\phi_0$ production, followed by $t\bar t\to b\bar bW^+W^-\to b\bar b \ell\bar\nu\bar\ell\nu$ and $\phi_0\to b\bar b$, at the HL-LHC with center-of-mass energy of $13.6$ TeV. The discriminative ability of each network determines how effectively it distinguishes between signal and background features, a metric quantified by the Receiver Operating Characteristic (ROC) curve. Enhanced discrimination performance is indicated by a higher true positive rate compared to the false positive rate in the ROC curve. To optimize the performance of each network, we individually adjust the cut on the ROC curve, aiming to boost the signal-to-background yield by calculating $\frac{S}{\sqrt{S+B}}$ at each bin of the ROC curves. Post-application of these cuts, the number of signal and background events is taken into account to compute the signal significance and ensuing limits.

\begin{figure}[!t]
    \centering
    \includegraphics[scale=0.33]{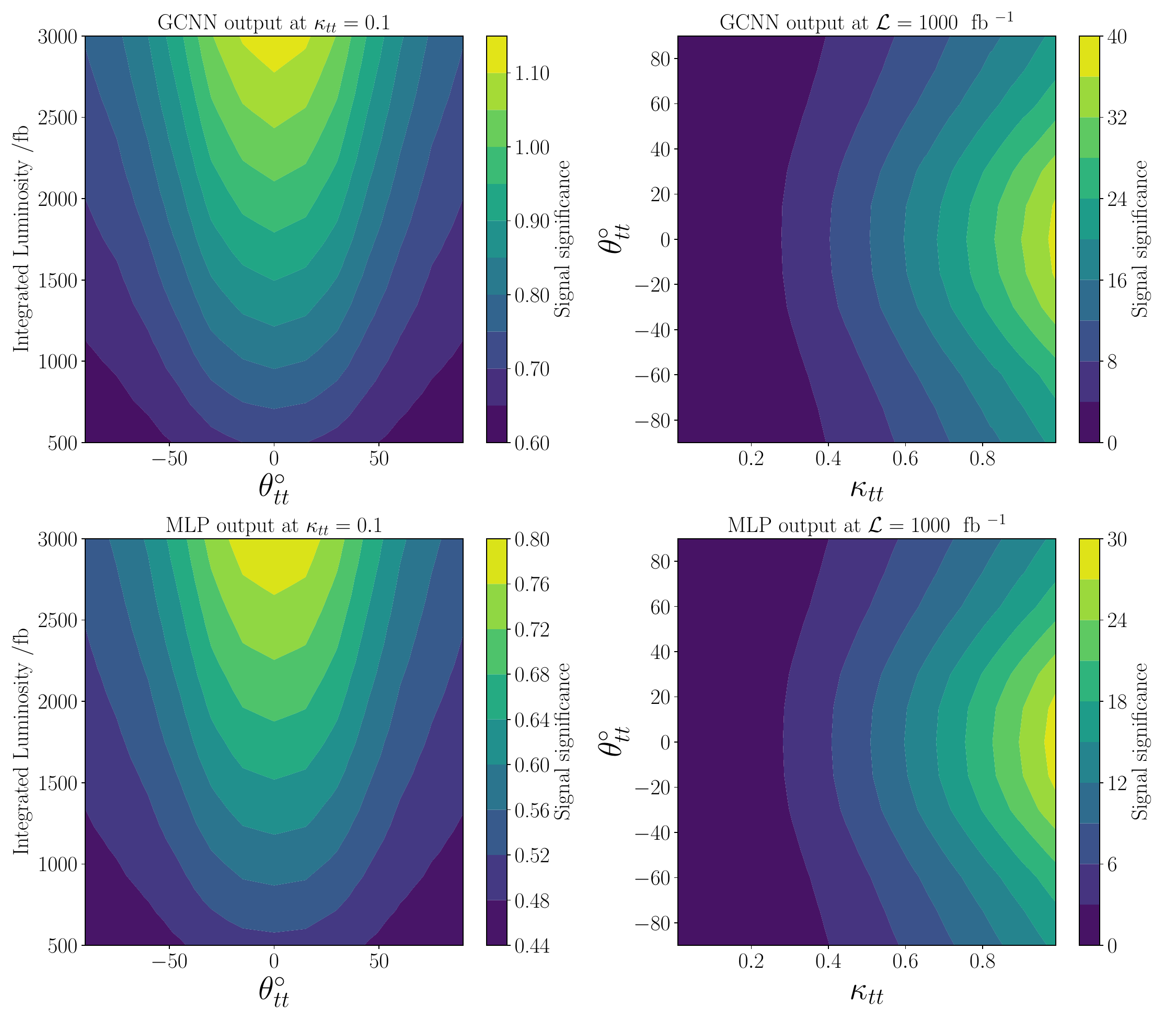}
    \caption{Left: contour plot for varying integrated luminosity and $\theta_{tt}$ at fixed value of $\kappa_{tt}=0.1$. Right: contour plot for varying $\theta_{tt}$ and $\kappa_{tt}$ at fixed integrated luminosity ${L}=1000 $ fb $^{-1}$. The contour colors represent the value of the signal significance. Finally, the upper raw represents the GNN results while the lower one represents the MLP results.}
    \label{fig:limits}
\end{figure}

This enables us to test the 
signal discovery hypothesis or setting an upper limit on the total cross section under the non-observation hypothesis. These can be determined by optimizing the signal-to-background cut on the DNN output, achieved using the following significance formula \cite{Cowan:2010js,LHCDarkMatterWorkingGroup:2018ufk,Antusch:2018bgr,Antusch:2020fyz}:
\begin{equation}
\sigma_{sys} = \left[ 2\left( (N_s+N_b)\ln\frac{(N_s+N_b)(N_b+\sigma^2_b)}{N_b^2+(N_s+N_b)\sigma^2_b} -\frac{N^2_b}{\sigma^2_b}\ln(1+\frac{\sigma^2_b N_s}{N_b(N_b+\sigma^2_b)}) \right) \right]^{1/2}\,,
\end{equation}
where $N_s$ and $N_b$ represent the counts of signal and background events, respectively, and $\sigma_b$ denotes the total uncertainty in the background events. Fig.  \ref{fig:limits} shows the signal significance on the $\kappa_{tt}$ for $\theta_{tt}$ values ranges from $-\frac{\pi}{2}$ to $\frac{\pi}{2}$ for the GNN (top) and MLP (bottom). Left plots display the contours for varying integrated luminosity and $\theta_{tt}$ at fixed value of $\kappa_{tt}=0.1$ while the right plots display the contours for varying $\theta_{tt}$ and $\kappa_{tt}$ at fixed integrated luminosity ${L}=1000 $ fb $^{-1}$. For all results the GNN shows an improvement over the MLP.

\begin{figure}[!t]
    \centering
    \includegraphics[scale=0.45]{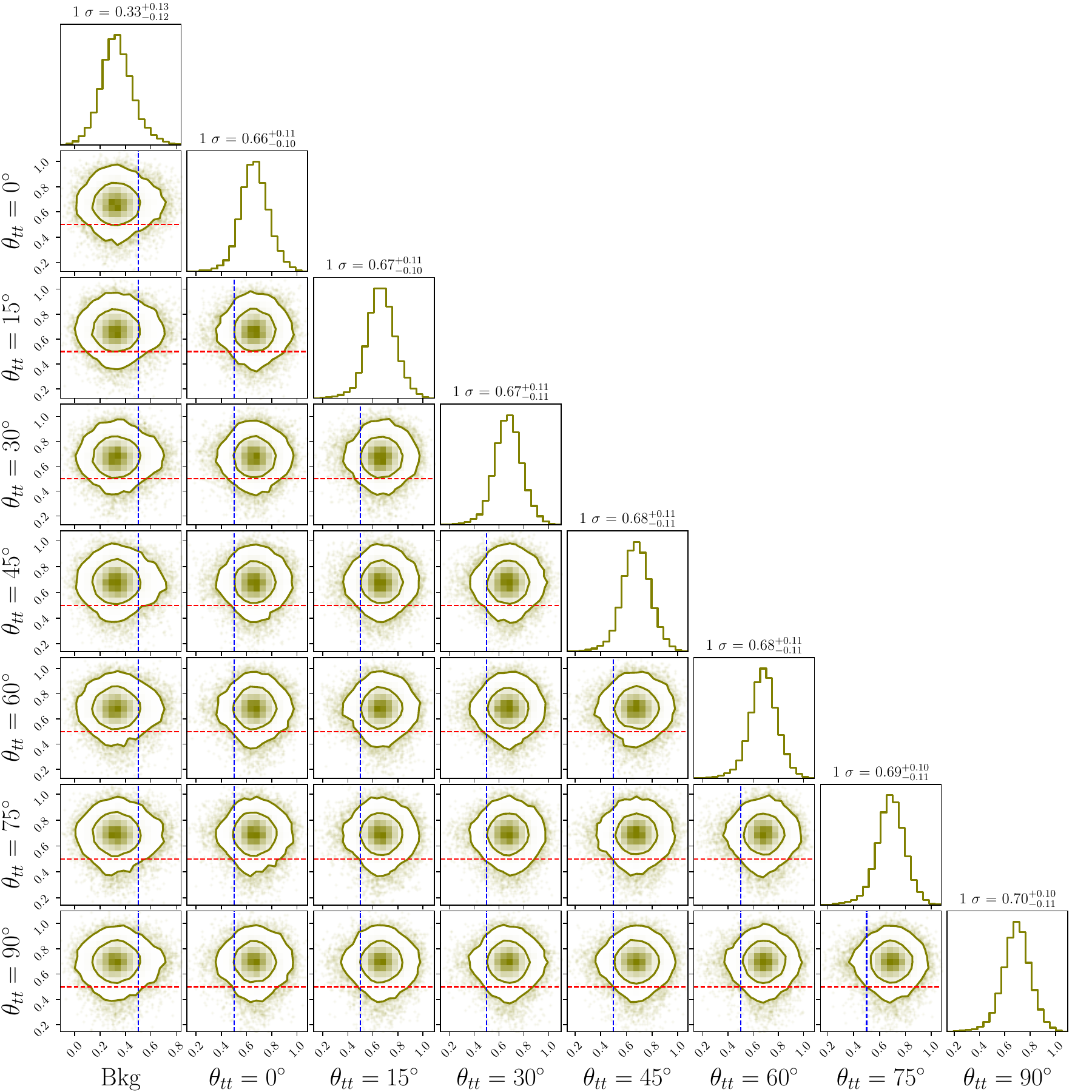}
    \caption{ Contour plots for $1$ ($68\%$ CL) and $2$ sigma ($95\%$ CL) of the MLP output when tested on different signal benchmark points using $30000$ test events. The model is trained on $\theta_{tt}=0^\circ,30^\circ,45^\circ,90^\circ$ and interpolates the results for $\theta_{tt}=15^\circ,60^\circ,75^\circ$. The $50\%$ quantile and $1\sigma$ error are written on top of each histogram. Events with MLP output near $1$ are considered as most likely signal events while those  with MLP output near $0$ are considered as most likely background events. Blue and red dashed lines indicates the $0.5$ value of the network output. For background correlation with all signal points, the upper left corner represents the true classified signal and background events while other corners represent the mis-identified rates.  For signal-to signal correlations the upper right corner represents the true classified events.}
    \label{fig:contour}
\end{figure}

As mentioned previously, the conditional network is used to interpolate the significance for new signal points with different $\theta_{tt}$ values. To test the network performance for the interpolation we test the correlation of the network output of the background and all tested signal events. Fig. \ref{fig:contour} shows the contour plots for $1$ ($68\%$ CL) and $2$ sigma ($95\%$ CL) of the MLP output when tested on different signal benchmark points using $30000$ test events\footnote{We opted to present the MLP results only, withe positive phase values, as the GNN output has similar response for interpolation.}.  The model is trained on $\theta_{tt}=0^\circ,\pm 30^\circ,\pm 45^\circ,\pm 90^\circ$ and interpolates the results for $\theta_{tt}=\pm 15^\circ,\pm 60^\circ,\pm 75^\circ$. The $50\%$ quantile and $1\sigma$ error is written on top of each histogram. Events with MLP output near $1$ is considered as most likely signal events, while those  with MLP output near $0$ are considered as most likely background events. Blue and red dashed lines indicates the $0.5$ value of the network output. For background correlation with all signal points, the upper left corner represents the true classified signal and background events while other corners represent the mis-identified rates.    For signal-to-signal correlations the upper right corner represents the true classified events. In short, it is clearly shown that the network is able to correctly interpolate to new points within $1\sigma$ level. Furthermore, this figure also indicates that conditional MLP has a good interpolation performance to new points within $1\sigma$ level.

\begin{figure}
    \centering
    \includegraphics[width=0.5\linewidth]{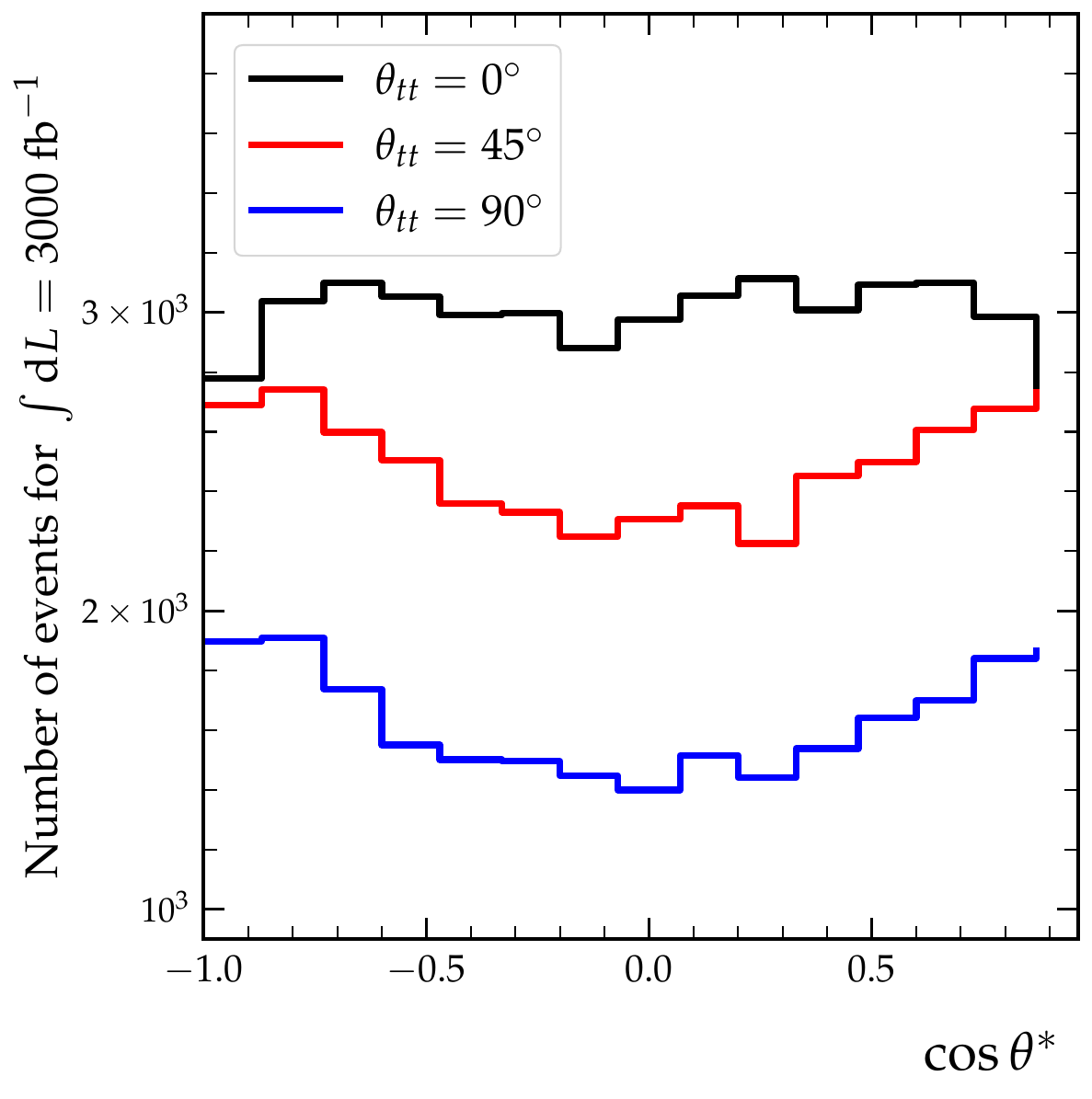}
    \vspace{-0.5cm}
    \caption{Distribution of $\cos\theta^*$, as defined in equation \eqref{eq:costhetaS}, in the signal region determined by the GCN training for $\theta_{tt} = 0^\circ$ (black), $\theta_{tt} = 45^\circ$ (red) and $\theta_{tt} = 90^\circ$ (blue). The distribution is obtained after tuning the cut on the GCN output probability to maximize the signal-to-background ratio. For the three cases, we add the QCD contribution post-GCN training, which is properly normalized. A $b$-tagging efficiency of $\epsilon_b = 70\%$ is assumed. (The $\cos\theta^*$ behavior for the pre-GCN cut showing a comparison between the pure ${\rm CP}$-odd, pure ${\rm CP}$-even and QCD cases is shown in the left-top panel of Fig. \ref{fig:distributions}.)}
    \label{fig:costhetas_postfit}
\end{figure}

\begin{figure}[!t]
    \centering
    \includegraphics[width=0.49\linewidth]{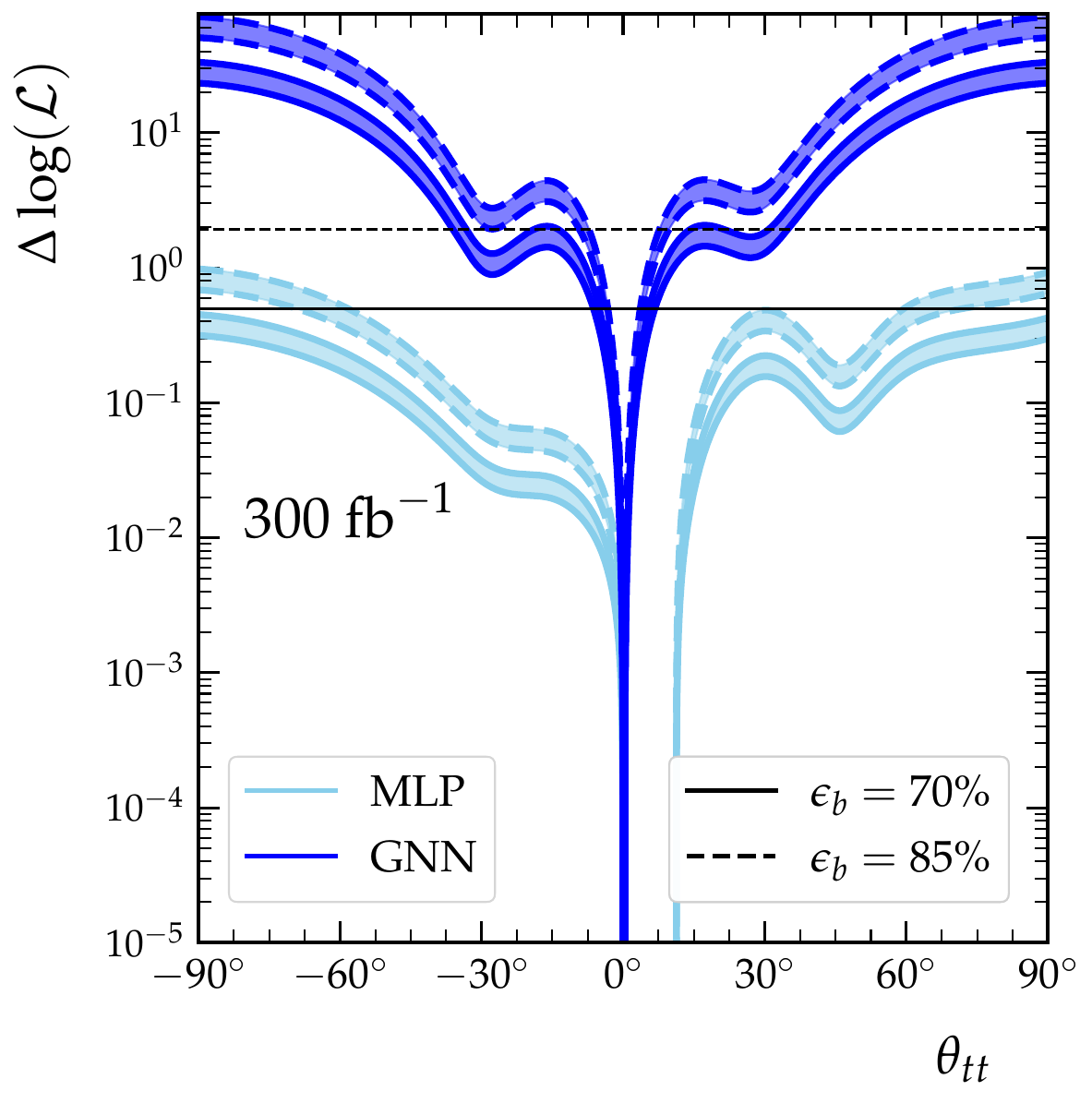}
    \includegraphics[width=0.49\linewidth]{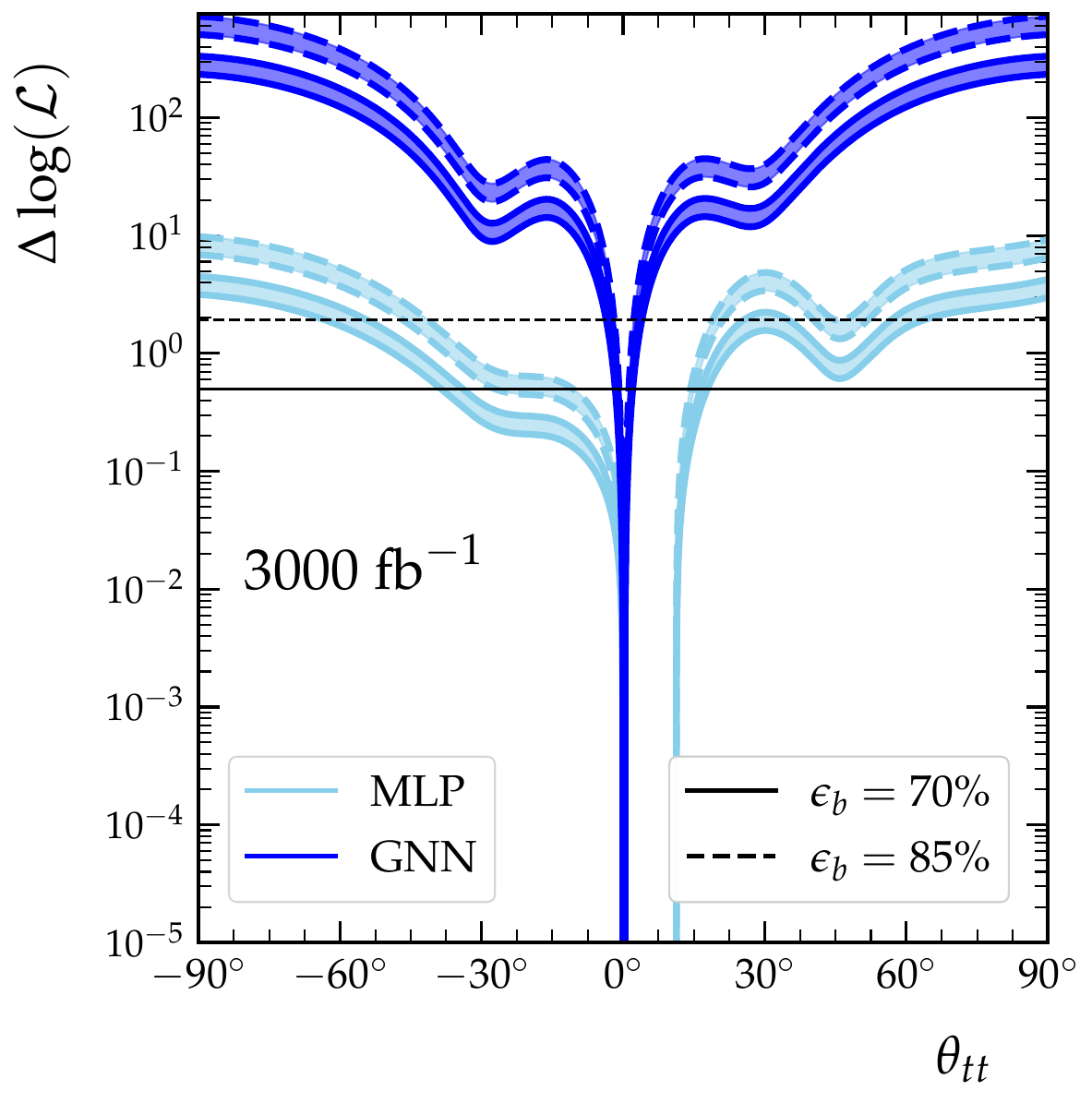}
    \caption{The log likelihood as a function of $\theta_{tt}$ for the two networks: MLP (cyan) and GNN (blue) of $300~{\rm fb}^{-1}$ (left) and $3000~{\rm fb}^{-1}$ (right). In each case we show the results for two nominal $b$-tagging efficiencies of $70\%$ (solid) and $85\%$ (dashed). The solid and dashed black lines correspond to the $68.4\%$ and $95.3\%$ confidence levels respectively. The dashed areas correspond to the variation of the likelihood due to uncertainties of $5\%$ and $20\%$. All the results are shown for $\kappa_{tt} = 1$.}
    \label{fig:LL:result}
\end{figure}

We close this section by showing the results of the shape analysis of the observables after the MLP and GNN optimization. Events passing the signal region definition, are used for the calculation of the Log-Likelihood function defined as 
\begin{eqnarray}
    \log({\cal L}) = - \sum_i \bigg[n_i \log\left(\frac{n_i}{\nu_i}\right) + \nu_i - n_i \bigg],
\end{eqnarray}
with $n_i$ and $\nu_i$ being the number of events for the expected ($\theta_{tt} = 0^\circ$ + QCD) and the alternative ($\theta_{tt} \neq 0^\circ$ + QCD) hypotheses respectively. The sum runs over all the bins and all the observables being used. For the MLP case, we have used four observables as input: $\Delta\phi_{\ell^+\ell^-}$, $\cos\varphi_{\ell\ell}$, $\Delta\phi_{\ell\ell}^{tt}$ and $\cos\theta^*$ while for the GNN case we have used three observables: $\cos\varphi_{\ell\ell}$, $\cos\theta_{\ell}^k$ and $\cos\theta^*$ (an example is shown in Fig. \ref{fig:costhetas_postfit}). It is found that these observables are the most sensitive to the ${\cal CP}$ nature and show the best feature importance in the training stage of the NN algorithms. The results for the binned log-likelihood are shown in Fig. \ref{fig:LL:result} for 300 ${\rm fb}^{-1}$ (left) and 3000 ${\rm fb}^{-1}$ (right) for the two networks being considered in our analysis. For $\kappa_{tt} = 1$, the GNN algorithm shows a superior performance as compared to the MLP algorithm as we can see that already 300 ${\rm fb}^{-1}$ is enough to probe $\theta_{tt}$ of order $20^\circ$. We summarize below the expected exclusions at 95.4$\%$ CL:

\begin{eqnarray}
\theta_{tt}^{\rm MLP} \in \begin{cases}
		[-60^\circ, 60^\circ] & {\rm for} \quad {\cal L} = 300~{\rm fb}^{-1}, \\
		[-45^\circ, 30^\circ] & {\rm for} \quad {\cal L} = 3000~{\rm fb}^{-1}. 
		\end{cases} \nonumber \\
\theta_{tt}^{\rm GNN} \in \begin{cases}
		[-10^\circ, 10^\circ] & {\rm for} \quad {\cal L} = 300~{\rm fb}^{-1}, \\
		[-5^\circ, 5^\circ] & {\rm for} \quad {\cal L} = 3000~{\rm fb}^{-1}, 
		\end{cases}
\end{eqnarray}
where we assumed $\epsilon_b = 85\%$. We note that the results for case of $\epsilon_b = 70\%$ are not extremely different as can be seen in Fig. \ref{fig:LL:result}. Moreover, GNN results surpass the combined analysis in \cite{Barman:2021yfh} in which the ${\cal CP}$  phase is excluded for $|\theta_{tt}| \ge 18^\circ$ at $\sqrt{s} = 14$ TeV and $\mathcal{L} = 3000~\rm fb^{-1}$, assuming $k_{tt}=1$. Moreover, a recent CMS study \cite{CMS:2022dbt} has set a limit on the $t\bar{t}H$ coupling parameters from combined searches of $tH$ + $t\bar{t}H$ in the multi-lepton, $ZZ$ and $\gamma\gamma$ channels employing the eXtreme Gradient Boosting ({XGBoost}) method and low- and high-level features at a center-of-mass energy of $13$ TeV and $138~$ fb$^{-1}$. Recasting their bounds in terms of our Lagrangian parameters and for $\kappa_{tt}$, we found that the CMS analysis excludes $|\theta_{tt}| \gtrsim 33^\circ$ at $68\%$ CL\footnote{A transformation from the CMS parametrization of the $t\bar{t}H$ coupling and our parametrization leads to the following relations:
\begin{equation}
\kappa_{tt} = \sqrt{\kappa_t^2 + \tilde{\kappa}_t^2}, \qquad \theta_{tt} = \arctan{\left(\frac{\tilde{\kappa}_t}{\kappa_t}\right)},   
\end{equation}
where $\kappa_t$ and $\tilde{\kappa}_t$ are the ${\cal CP}$-even and ${\cal CP}$-odd components of the $t\bar{t}H$ coupling.}. While the results of our informed GNN methods are superior to those of CMS limits, we emphasize that a direct comparison is not possible, as the CMS analysis rigorously accounts for systematic uncertainties, which are beyond the scope of our phenomenological study.

%%%%%%%%%%%%%%%%%%%%%%%%%%%%%%%%%%
\section{Conclusions}
\label{sec:summary}
%%%%%%%%%%%%%%%%%%%%%%%%%%%%%%%%%%%%%%%%
DL is currently the state-of-the-art approach in many ML applications to particle physics (amongst other disciplines), yet the evaluation and training of DL models is generally still quite time-consuming and altogether computationally expensive. The so-called conditional computation approach has been thus proposed to tackle such a problem, as it operates by selectively activating only parts of the concerned network at a time. We have thus embraced it here in two different implementations:  MLP and GNN.

Armed with such computational tools, we have chosen a particle physics problem which is particularly suited to these approaches, given the large multiplicity of the final state (having eight particles to start with), the necessity of reconstructing masses of intermediate objects (five of these) and, finally, the need of extracting subleading CPV effects from a large variety of kinematical   observables (nine of these). The target process was $pp\to t\bar t \phi_0$, where $\phi_0$ is a generic neutral Higgs boson, which is currently being targeted by the ATLAS and CMS collaborations for the purpose of measuring the Yukawa coupling between the top (anti)quark and the SM-like Higgs boson discovered in 2012 ($h_{\rm SM}$). In such a process, this is done in so-called `open production' (wherein the top (anti)quark is produced as a real object in the final state), so as to be able to compare it against the same coupling measured in so-called `closed production' (wherein the top (anti)quark is produced as a virtual object in the loop of gluon-gluon fusion). Possibly more importantly, the extraction of such a coupling in the former case offers the unique possibility of testing the ${\cal CP}$ properties of the interaction between $\phi_0$ and $t\bar t$, by exploiting correlations amongst the momenta of the decay products of both the top (anti)quark pair and Higgs boson. In particular, in our analysis, we have assumed leptonic decays of the $t\bar t$ system and $b\bar b$ ones of the $\phi_0$ state. Given that such ${\cal CP}$ properties can only be accessed through differential distributions (rather than inclusively at integrated cross section level), specifically, through their characteristic line-shapes, a significant number of events is necessary for this purpose, hence, as collider setup, we have chosen here the HL-LHC. 

Following a sophisticated MC analysis based on, again, state-of-the-art event generation down to the detector level, we have been able to prove the superiority of our (conditional) MLP and GNN approaches with respect to more traditional ones, wherein either a cut-and-count selection is solely exploited or else this is used in combination with more trivial ML informed methods.
In particular, assuming the projected energy and luminosity of the HL-LHC, we demonstrate that it is possible to achieve sensitivity to CP-violating phase values ranging from $0$ to $\frac{\pi}{2}$.
%In particular, assuming foreseen energy and luminosity of the HL-LHC, we have proven that one can establish sensitivity to all CPV phase values between $-\frac{\pi}{2}$ and $\frac{\pi}{2}$, including distinguishing between CP-even and -odd components in the signal sample.
We found that MLP and GNN approaches are viable alternatives to more traditional methods. Furthermore, we have also tensioned the MLP against the GNN implementation and found that the latter exhibits better performance than the former.
Finally, notice that, in order to unable validation of the results obtained here, our code and data are available on  GitHub at 
\href{https://github.com/AHamamd150/Conditional_GNN} {\tt https://github.com/AHamamd150/Conditional$_{-}$GNN}.

\section*{Acknowledgments}
%%%%%%%%%%%%%%%%%%%%%%%%%%%%%%%%%%%%%%%%%%%%%%%%%%%%%%%%%%%%
 AH is funded by the Grant Number 22H05113 from the ``Foundation of Machine Learning Physics'', the ``Grant in Aid for Transformative Research Areas''  22K03626 and the Grant-in-Aid for Scientific Research (C). The work of AJ is supported by the Institute for Basic Science (IBS) under Project Codes IBS-R018-D1 and IBS-R018-D3. SM is supported in part through the NExT Institute and the   STFC Consolidated Grant ST/L000296/1. WE is funded by the ErUM-WAVE project 05D2022 "ErUM-Wave: Antizipation 3-dimensionaler Wellenfelder", which is supported by the German Federal Ministry of Education and Research (BMBF).

%%%%%%%%%%%%%%%%%%%%%%%%%%%%%%%%%%%%%%%%%%%%%%%%%%%%%%%%%%%%%%
\appendix
\section{Verify the network performance with a toy example }
\label{app:A}
\begin{figure}[!h]
    \includegraphics[scale=0.3]{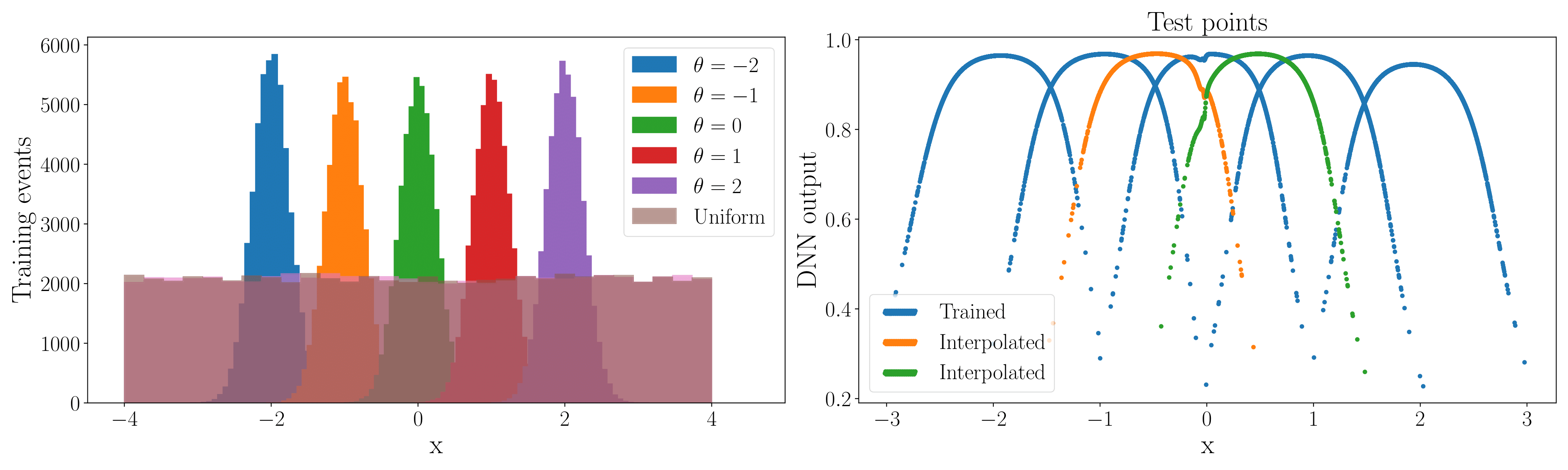}
    \vspace{-0.3cm}
    \caption{Left: training distributions in which the signal are represented by Gaussian and the background represented by a uniform distribution. Right: the network output  as a function of the value of the input feature $x$, for various choices of the input parameter $\theta$. Orange and green distributions represent the points that the network has not seen during the training with $\theta = -0.5, 0.5$, respectively. }
    \label{fig:toy}
\end{figure}

%%%------------------------------------------%%
Following the methodology outlined in \cite{Baldi:2016fzo} for conditional DNN, we validate our network architecture by replicating the illustrative toy example presented in the aforementioned paper. This example involves a simplified scenario featuring a single feature $x$ and a corresponding conditional parameter $\theta$. The input features are Gaussian distributions with mean values equal to $\theta$ and a standard deviation of $\sigma=2.5$. Specifically, we examine signal points corresponding to $\theta = -2, -1, 0, 1, 2$, while the background points follow a uniform distribution, as depicted in Figure \ref{fig:toy} (left).

In preparing the training dataset, we generate $500,000$ signal points by stacking equally sized features for each $\theta$ value. Since the primary objective of the network is to learn the distinguishing features of the signal events and interpolate between them, we incorporate smaller number of background points specifically  $150,000$  points, but with equal size of siganl and background for test. For the conditional input vector, $\theta$, we concatenate the five values corresponding to each $x$, resulting in a single vector of length $500,000$. For background events, we create a vector of random variables ranging between $-2$ and $2$. The labels assigned to signal points are $Y=1$, while background points are labeled as $Y=0$.

For this example, we consider a conditional MLP that has two input layers, one for the feature $x$ and one for the conditional parameter $\theta$. Input layer of the feature $x$ is followed by three FC layers with number of neurons $300,300,100$ and ReLU activation function; while the second input layer is followed by a linear layer with $100$ neurons and no activation function. The two layers with $100$ neurons are concatenated and passed to an output layer with one neuron and sigmoid activation function. The model is trained with $5$ epochs and batch of size $500$ points. We use a mean squared error function and Adam optimizer with learning rate of $10^{-4}$.

The network attains a training accuracy of $95.8\%$ and a test accuracy of $95.7\%$. The network output is shown in Figure \ref{fig:toy} (right plot) in blue. To evaluate its performance further, we introduce new points with $\theta$ values not included in the training set, specifically $-0.5$ and $0.5$, represented by the orange and green distributions, respectively. Remarkably, the network demonstrates its capability to interpolate to these novel $\theta$ values, despite not being explicitly trained on them.
%%%%%%%%%%%%%%%%%%%%%%%%%%%%%%%%%%%%%%%%%%%%%%%%%%%%%%%%%%

%%%%%%%%%%%%%%%%%%%%%%%%%%%%
\bibliographystyle{JHEP}
\bibliography{main}

%%%%%%%%%%%%%%%%%%%%%%%%%%%%%%%%
\end{document}